\documentclass[11pt]{article}

\usepackage{sectsty}
\sectionfont{\Large}
\subsectionfont{\large}
\usepackage{hyperref}
\hypersetup{colorlinks,urlcolor=blue,citecolor=black}
\usepackage{authblk}
\usepackage{multirow,array}

\usepackage{xcolor}
\usepackage{graphicx}
\usepackage[left=1in,right=1in,top=1.2in,bottom=1.2in,letterpaper]{geometry}
\usepackage{times}
\usepackage{caption,setspace}
\usepackage{lscape} 

\usepackage{ulem}

\usepackage[authoryear]{natbib}

\usepackage{amsmath, amsthm, amssymb, amsfonts}

\newtheorem{thm}{Theorem}

\numberwithin{equation}{section}

\begin{document}

    \title{Two-phase analysis and study design for survival models with error-prone exposures}

	\author[1]{Kyunghee Han\footnote{Email: \href{mailto:Kyunghee}{kyunghee.stat@gmail.com}}}
	\author[2]{Thomas Lumley}
	\author[3]{Bryan E. Shepherd}
    	\author[1]{Pamela A. Shaw}
    	\affil[1]{Department of Biostatistics, Epidemiology, and Informatics, University of Pennsylvania, USA}
    	\affil[2]{Department of Statistics, University of Auckland, New Zealand}
    	\affil[3]{Department of Biostatistics, Vanderbilt University, USA}
    	\date{
    	}

\maketitle

\begin{abstract}
Increasingly, medical research is dependent on data collected for non-research purposes, such as electronic health records data (EHR). EHR data and other large databases can be prone to measurement error in key exposures, and unadjusted analyses of error-prone data can bias study results. Validating a subset of records is a cost-effective way of gaining information on the error structure, which in turn can be used to adjust analyses for this error and improve inference. We extend the mean score method for the two-phase analysis of discrete-time survival models, which uses the unvalidated covariates as auxiliary variables that act as surrogates for the unobserved true exposures. This method relies on a two-phase sampling design and an estimation approach that preserves the consistency of complete case regression parameter estimates in the validated subset, with increased precision leveraged from the auxiliary data. Furthermore, we develop optimal sampling strategies which minimize the variance of the mean score estimator for a target exposure under a fixed cost constraint. We consider the setting where an internal pilot is necessary for the optimal design so that the phase two sample is split into a pilot and an adaptive optimal sample. Through simulations and data example, we evaluate efficiency gains of the mean score estimator using the derived optimal validation design compared to balanced and simple random sampling for the phase two sample. We also empirically explore efficiency gains that the proposed discrete optimal design can provide for the Cox proportional hazards model in the setting of a continuous-time survival outcome. 
\end{abstract}


\section{Introduction}

Error-prone exposures are common in many epidemiological settings, such as clinical studies relying on electronic health records, medical claims data, or large observational cohort studies where a gold standard measure was not collected on all subjects. In some cases, a validation subset is available, in which observations without error can be obtained on a subset of subjects and used to inform measurement error correction methods. Two-phase sampling has been widely used for a number of settings in clinical and epidemiological studies with budgetary constraints. The first sampling phase includes readily available data (e.g., electronic health records data) on all study subjects, and the second phase includes additional information on a subsample of records (e.g., extensive chart validation of a key exposure variable). 

The efficiency of two-phase sampling can vary substantially based on the selection of the second phase sample. For logistic regression, \cite{breslow1999design} showed stratification of the phase two sample on the outcome and covariates with equal numbers per stratum performed well and better than stratifying on the outcome or covariates alone. When the outcome and error-prone exposure are discrete, the mean score method \citep{reilly1995mean, reilly1996optimal} can be used to derive regression parameter estimates that have been corrected for measurement error and can improve efficiency over the complete case analysis by incorporating information from auxiliary data. Further, these authors developed an optimal sampling strategy for the mean score estimator to find the proportion allocated into each outcome-exposure stratum of the validation sample that can minimize cost for a fixed variance or minimize the variance of a target regression parameter for a fixed validation subset size, studying the performance of this method for a binary outcome and covariate.  \cite{mcisaac2014response} compared response-dependent two-phase sampling designs for the setting of a binary outcome when both the true and auxiliary covariate were also binary. They found that the mean score estimator was an efficient approach, even when the model used to derive the optimal design was misspecified. They also found that the mean score optimal design improved the efficiency of other estimation approaches, such as maximum likelihood. 

For survival outcomes, however, the focus of most of the previous work on two phase sampling has been on estimation and hypothesis testing and not design. \cite{lawless2018two} reviews two-phase estimators for outcome dependent sampling and failure time data, and mentions that efficiency can be gained by sampling extremes of the outcome those with early events and late censoring. \cite{tao2019optimal} develop an optimal two-phase sampling designs for a general regression model framework, but these designs are only optimal under the null assumption that the regression coefficient for the expensive/mis-measured exposure of interest is zero.  While this framework in principle applies to the Cox proportional hazards regression model, these designs require estimation of several nuisance parameters related to the conditional distribution of expensive covariate given the surrogate, as well as a potentially infinite dimensional nuisance parameter related to the log-partial likelihood; and their performance have not yet been studied for survival outcomes. 

In this study, we consider an extension of the mean score method to handle survival outcomes, in which error prone exposures are treated as auxiliary variables. The mean score approach is appealing because it is intuitive and offers a straightforward method to implement an optimal design \citep{whittemore1997multi}. In order to take advantage of the mean score approach, we will first consider a discrete-time survival model, but will also consider how the derived optimal design can be advantageous for a continuous time analysis. Discrete survival data are natural for settings where the occurrence of an event is monitored periodically and occur frequently in clinical studies where there is routine follow-up at fixed intervals.  We develop an application of the mean score method to the discrete proportional hazards model. We will also extend the work of \cite{reilly1996optimal} to derive an optimal sampling design for the validation subset, which minimizes the variance of the regression parameter estimation for a given size of the validation subset. This approach requires an estimate of several nuisance parameters in the absence of external estimates, which can be estimated with internal pilot data. We consider a multi-wave sampling strategy that in the first wave obtains  a pilot phase two sample to estimate the parameters necessary to derive the optimal design and then for the second wave adopts an adaptive sampling strategy for the remaining phase two subjects to achieve the optimal allocation. \cite{mcisaac2015adaptive} considered a similar approach for binary outcomes.  

We compare the relative efficiency of our mean score estimator under simple random sampling, balanced sampling and the proposed optimal design. We also examine the efficiency gains of the mean score approach over the complete case estimator of the validated data. The proposed method is further illustrated with data from the National Wilms Tumor Study (NWTS). In the NWTS, the true validated exposure was measured on everyone, allowing the evaluation of different two-phase sampling strategies on the precision of the final study estimates in this applied setting. In the context of this example, which had a continuous survival outcome, we study the efficiency gains of using the proposed optimal design of the discretized outcome for the usual continuous-time analysis.  We also investigate how different allocations of the pilot sample affect the efficiency gains of the mean score estimator, depending on how individuals with censored versus observed outcomes are sampled. Finally, we provide some concluding remarks on the advantages of the mean score estimator in this setting and discuss directions for future work.

	\section{The mean score method for discrete-time survival models} \label{setup}

	\subsection{Setup and notation for the discrete-time model} \label{subsec-notation}
	
Let $T $ be a discrete random variable. Denote the $j$-th discrete value of $T$ by $t_j$ and write $\lambda_{0j} = \lambda_0(t_j)$, where $\lambda_0$ is the baseline hazard function for $T$ defined by $\lambda_0(t_j) = P(T = t_j | T \geq t_j )$ for $j \in J$, where $J$ is the index set for the discrete times with positive mass. We assume the time to event response $T$ is associated with a $d$-dimensional covariate vector $\mathbf{X}=(X_1, \ldots, X_d)^\top$ such that the conditional hazard function $\lambda( t| \mathbf{x}) = P(T = t | T \geq t ,\, \mathbf{X}=\mathbf{x})$ is given by 
\begin{align}
	g\big(\lambda(t | \mathbf{x})\big) =g\big( \lambda_0(t) \big) \exp(\boldsymbol{\beta}^\top \mathbf{x}) \quad (t \in \mathcal{T}) \label{phz-model}
\end{align}
for some coefficient vector $\boldsymbol{\beta} = (\beta_1, \ldots, \beta_d)^\top$, where $g:[0,1] \rightarrow \mathbf{R}$ is a monotone transformation and $\mathcal{T} = \{t_j : j \in J \}$. For example, the odds transformation $g_1(u) = \frac{u}{1-u}$ yields the logit hazard model, where $\textrm{logit} \big( \lambda_j(\mathbf{x}) \big) = \alpha_j + \boldsymbol{\beta}^\top \boldsymbol{x}$ for $\alpha_j = \textrm{logit}(\lambda_{0j})$ and $\lambda_j(\mathbf{x}) = \lambda(t_j | \mathbf{x})$. It follows that the likelihood function under the logit hazard model has the same representation with the logistic regression model such that the number of events at a time $t_j$ represents binomial outcomes with probability $\lambda_j(\mathbf{x})$ for each $j \in J$ \citep{meier2003discrete}. The complementary log transformation $g_2(u) = -\log(1-u)$ gives a proportional log-survival model $\log \big(S_j( \mathbf{x})\big) = \log (S_{0j}) \exp(\boldsymbol{\beta}^{\top} \mathbf{x})$, where $S_j(\mathbf{x}) = P(T \geq t_j | \mathbf{X}=\mathbf{x})$ and $S_{0j}=S_0(t_j)$ for the baseline survival function $S_0$ for $T$. \cite{kalbfleisch2011statistical} provide further details.

We consider that $T$ may be subject to random censoring prior to the finite maximum follow-up time of $\tau < \infty$. For a random censoring time $C$, independent of $T$, let $Y = \min\{ T, C \}$ be the observed censored survival time and $\Delta = \mathbb{I} (T \leq C)$ be the event indicator. We assume that $\mathcal{X}_N = \{ (Y_i, \Delta_i, \mathbf{X}_i) : 1 \leq i \leq N \}$ are independent and identically distributed and will not be completely observed on all subjects as $N$-random copies of $(Y,\Delta,\mathbf{X})$. Instead,  $\mathcal{X}_{\textrm{I},N} = \{ (Y_i, \Delta_i, \mathbf{Z}_i):  1 \leq i \leq N \}$ are the data available for all subjects in the first phase of a study, where $\mathbf{Z}=(Z_1, \ldots, Z_q)^\top$ are discrete surrogates or auxiliary variables associated with $\mathbf{X}$. Complete data for the likelihood are observed on the phase two sample, denoted by $\mathcal{X}_{\textrm{II},n} = \{ (Y_i, \Delta_i, \mathbf{X}_i) : i \in \mathcal{I} \}$, where $\mathcal{I}$ is a subset of $\{1, \ldots, N\}$ having the cardinality of $n$. For example, the auxiliary variables $\mathbf{Z}$ might be error-prone discrete measures of $\mathbf{X}$ while $\mathbf{X}$ may or may not be discrete. We also allow for settings where some components of $\mathbf{X}$ are available on all subjects, such as sex or other demographic information ascertained in the first phase of the study. For this setting, we may introduce a slight abuse of notation writing $\mathbf{X} = (\mathbf{X}^C, \mathbf{X}^I)$ and $\mathbf{Z} = (\mathbf{X}^C, \mathbf{A})$, where $\mathbf{X}^{C}$ are the components of $\mathbf{X}_i$ observed on everyone, $\mathbf{X}^{I}$ are the incomplete components of $\mathbf{X}$ not observed at the phase one, and $\mathbf{A}$ indicates a generic notation for auxiliary variables in the first phase of the study. In this case, $\mathcal{I}$ denotes the set of subject indices for which the true covariates $\mathbf{X}$ are fully observed, with $\mathbf X^{I}$ sampled in the second phase of the study, and the auxiliary phase one variables are limited to $\mathbf{A}$ \citep{reilly1995mean, mcisaac2015adaptive}. 
 
 Unlike previous literature \citep{lawless1999semiparametric, chatterjee2003pseudoscore, chatterjee2007semiparametric, tao2017efficient}, we note that the model \eqref{phz-model} does not include auxiliary variables as predictors, but rather $\mathbf{Z}$ is a surrogate by the Prentice criterion \citep{prentice1989surrogate}. That means the likelihood function   $L_1(\boldsymbol{\theta}; Y, \Delta, \mathbf{X}, \mathbf{Z})$ equals  $L_1(\boldsymbol{\theta} ; Y, \Delta, \mathbf{X})$, where $\boldsymbol{\theta}$ is the collection of parameters involved in \eqref{phz-model}.
Thus, the complete information for $\boldsymbol{\theta}$ is carried by $(Y, \Delta, \mathbf{X})$, while $ \mathrm{E} \{ \log L_1(\boldsymbol{\theta}; Y, \Delta, \mathbf{X}) \,|\, \allowbreak Y, \Delta, \mathbf{Z} \}$ may represent extra information of $(Y, \Delta, \mathbf{Z})$ compared to likelihood-based inference using only complete case data for $(Y,\Delta,\mathbf{X})$. In this paper, we consider the mean score method, which is valid when the validation subset $\mathcal{I}$ for the phase two sample is randomly chosen from the full cohort given information obtained from the first phase of a study. Therefore, we use the log-likelihood from all available observations written by
\begin{align}
	\sum_{i \in \mathcal{I}}  \log L_1(\boldsymbol{\theta} ; Y_i, \Delta_i, \mathbf{X}_i) + \sum_{i \in \mathcal{I}^c} \int \log L_1(\boldsymbol{\theta}; Y_i, \Delta_i, \mathbf{x}) h(\mathbf{x} |  \mathbf{Z}_i) \, \mathrm{d}\mathbf{x}, \label{likelihood}
\end{align}
where $h(\mathbf{x} | \mathbf{z})$ denotes the conditional density function of $\mathbf{X}$ given $\mathbf{Z}=\mathbf{z}$ and $\mathcal{I}^c = \{1, \ldots, N\} \setminus \mathcal{I}$ indicates a set of $(N-n)$ indices for individuals whose complete covariates are not available. For an individual $i$ and time $j$, we define the observed event indicator ${D}_{ij} = \mathbb{I}(Y_i = t_j, \,\Delta_i = 1)$ and denote subject $i$'s censored survival time index by $J(i) = \arg\{ j \in J : Y_i = t_j \}$ for $1 \leq i \leq n$. Thus,  $D_{ik} = 0$ for all $k < J(i)$. Then, it follows that the log-likelihood function given $(Y_i, \Delta_i, \mathbf{X}_i)$ can be written by
 \begin{align}
	\log L_1(\boldsymbol{\theta}; Y_i, \Delta_i, \mathbf{X}_i) 
		=  \sum_{j=1}^{J(i)} \Big[ {D}_{ij}  \log \Big(\frac{\lambda_j(\mathbf{X}_i)}{1-\lambda_j(\mathbf{X}_i)} \Big) + \log \big( 1- \lambda_j(\mathbf{X}_i) \big) \Big], \label{log-linear}
\end{align}
for each $1 \leq i \leq  n$. In the above equation, we used the fact that $L_1(\boldsymbol{\theta} ; Y, \Delta, \mathbf{X}) = S(Y| \mathbf{X}) \lambda(Y | \mathbf{X})^{\Delta} (1-\lambda(Y | \mathbf{X}))^{1-\Delta}$
and the conditional survival function $S_j(\mathbf{x}) = P(T \geq t_j | \mathbf{X}=\mathbf{x})$ was calculated by $\prod_{k=1}^{j-1} (1-\lambda_k(\mathbf{x}))$ for $j \geq 2$,
together with $S_1(\mathbf{x}) = 1$ by definition. 

\subsection{The mean score method} \label{subsec-mean score}

We apply the mean score method \citep{reilly1995mean, reilly1996optimal} to the conditional hazard model \eqref{phz-model} when auxiliary variables are discrete. Employing the Expectation-Maximization (EM) technique \citep{dempster1977maximum} with \eqref{likelihood} and \eqref{log-linear}, we may find the maximum likelihood estimator of $\boldsymbol{\theta}$. However, $h(\mathbf{x} | \mathbf{z})$ is generally unknown and a parametric approach for the estimation of $h(\mathbf{x} | \mathbf{z})$ may result in inconsistent inference of $\boldsymbol{\theta}$ in likelihood-based methods. \cite{lawless1999semiparametric} introduced a semi-parametric method, estimating the conditional density function $h(\mathbf{x} | \mathbf{z})$ nonparametrically, such that the integration in \eqref{likelihood} is replaced with a single-step approximation \citep{mcisaac2014response}. Following the mean score approach, for those not in the phase two subset, we can replace the unobserved score contribution based on $\mathbf{X}$ with its expected value based on the observed phase one data. By replacing the expected value to an empirical mean, the semi-parametric estimation of $\boldsymbol{\theta}$ can then be achieved by maximizing an inverse probability weighted log-likelihood
\begin{align}
	Q_{N}(\boldsymbol{\theta}) 
		&= \sum_{i \in \mathcal{I}} \sum_{j=1}^{J(i)} \hat{\pi}(Y_i, \Delta_i, \mathbf{Z}_i)^{-1} \Big[ {D}_{ij}  \log \Big(\frac{\lambda_j(\mathbf{X}_i)}{1-\lambda_j(\mathbf{X}_i)} \Big) + \log \big( 1- \lambda_j(\mathbf{X}_i) \big) \Big], \label{em-re}
\end{align}
where $\hat{\pi}(Y_i, \Delta_i, \mathbf{Z}_i)$ is an empirical estimate of ${\pi}(Y_i, \Delta_i, \mathbf{Z}_i)$, the sampling probability of the $i$-th individual selected into the validation subset, which can be consistently estimated by $n(Y_i, \Delta_{i}, \mathbf{Z}_i)/N(Y_i, \Delta_{i}, \mathbf{Z}_i)$ as $n$ and $N$ increase. Here, $n(y,\delta,\mathbf{z})$ is the number of subjects in $\mathcal{I}$ who have the same observations with $(y,\delta,\mathbf{z})$ in the first phase study; $N(y,\delta,\mathbf{z})$ is defined similarly to $n(y,\delta,\mathbf{z})$ with replacement of the index set $\mathcal{I}$ with $\{1, \ldots, N \}$ of the full sample. 

We note that \cite{flanders1991analytic} also proposed a similar estimating equation to \eqref{em-re} based on the pseudo-likelihood method, which coincides to the filling in method introduced by \cite{vach1991biased} when $\mathbf{X}$ is categorical. However, depending on the choice of transformation $g$ in \eqref{phz-model}, different forms of score equations follow from the above weighted log-likelihood \eqref{em-re}. In Supplementary Material Section \ref{technical-details}, we provide the detailed forms of the mean score equations and the associated Hessian matrices when the logit transformation $g_1$ and the complementary log transformation $g_2$ are used.

	\subsection{Connection to the Cox model for a continuous-time outcome} \label{subsec-cox}

Our expectation that the optimal design for our discrete time proportional hazards model will also be advantageous for the continuous time Cox model is based on the connection between the parameters in these two models. For this, we briefly review this connection. For further discussion, see \cite{kalbfleisch2011statistical}. 

From the log-likelihood \eqref{log-linear}, we note that the logit hazard model is the canonical form of the discrete-time survival model \eqref{phz-model} under the odds transformation $g_1(u) = \frac{u}{1-u}$, such that $\textrm{logit}(\lambda_j(\mathbf{x}))  =  \alpha_j+ \boldsymbol{\beta}^\top \mathbf{x}$, where $\alpha_j = \textrm{logit}(\lambda_{0j})$ is the logit transformation of the baseline hazard. However, we also note that any model with an arbitrary monotone transformation in \eqref{phz-model} leads to a reparameterization of the logit hazard model. For example, suppose the complementary log transformation $g_2(u) = -\log(1-u)$ defines the true survival model. Then it can be easily seen that $\textrm{logit}(\lambda_j(\mathbf{x}))  =  \exp\big(e^{\alpha_j+ \boldsymbol{\beta}^\top \mathbf{x}}\big) -1$, where $\alpha_j = \log(-\log(1-\lambda_{0j}))$ is the complementary log-log transformation of the baseline hazard, and the application of chain rules to \eqref{log-linear} is followed by likelihood-based estimation of $\boldsymbol{\theta} = (\boldsymbol{\alpha}, \boldsymbol{\beta})$, which is also equivalent to reparameterization of the logit hazard model. 

In particular, if usual continuous-time Cox proportional hazards are grouped into discrete disjoint intervals, this will lead to the conditional hazard model \eqref{phz-model} for discrete-time outcomes with the complementary log transformation. To be specific, let $\lambda^C(t | \mathbf{x})$ be the conditional hazard function in the Cox model for the continuous-time survival outcome such that $\lambda^C(t | \mathbf{x}) = \lambda_0^C(t) \exp(\boldsymbol{\beta}^\top \mathbf{x})$, 
where $\lambda_0^C$ is the associated baseline hazard function. 
Suppose we are only able to observe survival outcomes censored by disjoint intervals $(t_{j-1}, t_j]$ with $t_0\equiv0$, so that statistical inference on conditional hazards by each interval is a goal of the study. Then, we may denote the conditional hazards on the $j$-th interval by 
\begin{align}
	\lambda_j(\mathbf{x}) = P(T \leq t _j \,|\, T > t_{j-1}, \mathbf{X} = \mathbf{x} ). \label{cum-hazard}
\end{align}
for each $j \in J$. Since the conditional survival function $S^C(t | \mathbf{x}) = P(T > t \,|\, \mathbf{X} = \mathbf{x} )$ is equivalent to $\exp\big(- \int_0^t \lambda^C(s | \mathbf{x}) \, \mathrm{d}s \big)$ in the Cox model, we note that $ -\log \big( 1 - \lambda_j(\mathbf{x}) \big) = \int_{t_{j-1}}^{t_j} \lambda^C(s | \mathbf{x}) \, \mathrm{d}s $ represents the cumulative conditional hazard on the interval $[t_{j-1}, t_j)$ and, under the proportional hazards assumption, it can be shown that
\begin{align}
	-\log\big( 1-\lambda_j(\mathbf{x}) \big) = -\log\big( 1-\lambda_{0j} \big) \exp(\boldsymbol{\beta}^\top \mathbf{x}) \quad (j \in J), \label{cloglog}
\end{align}
where $\lambda_{0j} = 1 - \exp\big( - \int_{t_{j-1}}^{t_j} \lambda_0^C(s) \, \mathrm{d}s \big)$ is the associated cumulative baseline hazard on $[t_{j-1}, t_j)$. Thus, the cumulative hazard model \eqref{cloglog} is directly connected with the discrete-time survival model \eqref{phz-model} under the complementary log transformation, and the two models have the same regression coefficient $\boldsymbol{\beta}$.

	\section{Adaptive sampling design for optimal estimation} \label{sec-optsamp}

As introduced by \cite{reilly1995mean} and \cite{reilly1996optimal}, we consider a sampling design that asymptotically minimizes the variance of parameter estimates of interest under constraints that the validation sample size is fixed. In Theorem \ref{asymp-normal}, we establish that the asymptotic variance of the mean score estimator depends on the sampling probability for the phase two validation subset.  

\begin{thm} \label{asymp-normal}
For each $j \in J$, let $\alpha_j = g(\lambda_{0j}) \in \mathbf{R}$ be transformation of baseline hazards in \eqref{phz-model}. Suppose that the censoring time is bounded, that is $P(C \leq \tau)=1$ for some fixed constant $\tau > 0$, and that the conditional hazard functions $\lambda_j(\mathbf{x}) = \lambda(t_j | \mathbf{x})$ are bounded away from 0 and 1 for all $\mathbf{x} \in \mathbf{R}^d$. Under the regularity conditions in Supplementary Material Section \ref{conditions}, the mean score estimator $\hat{\boldsymbol{\theta}}$ of $\boldsymbol{\theta} = (\boldsymbol{\alpha}, \boldsymbol{\beta})$ solving the score equation of \eqref{em-re} is asymptotically normal such that $N^{1/2} \big( \hat{\boldsymbol{\theta}} - \boldsymbol{\theta} \big) \stackrel{\mathrm{d}}{\rightarrow} N\big( \boldsymbol{0}, \Sigma \big)$ as $N \rightarrow \infty$, where $\Sigma = I_V^{-1} + I_V^{-1} \Omega I_V^{-1}$ with $I_V = - \mathrm{E} \big[\frac{\partial^2}{\partial\boldsymbol{\theta}\partial\boldsymbol{\theta}^\top} \log L_1(\boldsymbol{\theta}; Y, \Delta, \mathbf{X}) \big]$ and $\Omega = \mathrm{E} \big[ \{\pi(Y,\Delta,\mathbf{Z})^{-1}-1\} \mathrm{Var}\big(U_1(\boldsymbol{\theta}) | Y, \Delta, \mathbf{Z} \big) \big] $ for the score function $U_1(\boldsymbol{\theta}) = \frac{\partial}{\partial\boldsymbol{\theta}} \log L_1(\boldsymbol{\theta}; Y, \Delta, \mathbf{X})$.
\end{thm}

Suppose that we fix the sample probability for the validation subset by $\pi_V = \mathrm{E} [\pi(Y,\Delta,\mathbf{Z})]$, or empirically $n/N$. Then, the mean score estimator of $\theta_k$, the $k$-th component of $\boldsymbol{\theta}$, is asymptotically efficient when the validation sampling probability $\pi(Y,\Delta,\mathbf{Z})$ is proportional to $\big\{ I_V^{-1} \mathrm{Var}\big(U_1(\boldsymbol{\theta}) | y, \delta, \mathbf{z} \big)  I_V^{-1} \big\}_{[k,k]}^{1/2}$, so that we empirically assign the optimal sampling size for each $(y,\delta,\mathbf{z})$-stratum
\begin{align}
	{n}^{\textrm{Opt}}(y,\delta,\mathbf{z}) \propto  N(y,\delta,\mathbf{z}) \big( {I}_V^{-1} {\mathrm{Var}}\big(U_1(\boldsymbol{\theta}) | y, \delta, \mathbf{z} \big) {I}_V^{-1} \big)_{[k,k]}^{1/2} \label{optimal-design}
\end{align}  
satisfying $n=\sum_{(y,\delta,\mathbf{z})} {n}^{\textrm{Opt}}(y,\delta,\mathbf{z})$, where $M_{[j,k]}$ denotes the $(j,k)$-element of a matrix $M$. We note that \eqref{optimal-design} can be viewed as Neyman allocation \citep{neyman1934two} maximizing the survey precision in the stratified sampling based on the phase one information. Such an optimal design can be also obtained by the Lagrangian multiplier method to minimize $\Sigma_{[k,k]}$ in Theorem \ref{asymp-normal}, which equivalently minimizes the variance of the target parameter $\theta_k$ with respect to $\pi(y, \delta, \mathbf{z})$ under the constraint of a fixed-validation rate $\pi_V = \mathrm{E} [\pi(Y,\Delta,\mathbf{Z})]$, or empirically $n= \sum_{(y,\delta,\mathbf{z})} \pi(y,\delta,\mathbf{z}) N(y,\delta,\mathbf{z})$. 

Note that the optimal sampling design depends on external information about population structure, namely $I_V$ and $\mathrm{Var}\big(U_1(\boldsymbol{\theta}) | y, \delta, \mathbf{z} \big)$, which are usually unknown.  \cite{mcisaac2015adaptive} introduced an adaptive procedure for multi-phase analyses such that one first draws a pilot sample for validation and then adaptively draws an additional validation set, write $\mathcal{X}_{\textrm{II},n}^\textrm{Pilot} = \{ (Y_i, \Delta_i, \mathbf{X}_i) : i \in \mathcal{I}^{\textrm{Pilot}} \}$ and $\mathcal{X}_{\textrm{II},n}^\textrm{Adapt}= \{ (Y_i, \Delta_i, \mathbf{X}_i) : i \in \mathcal{I}^{\textrm{Adapt}} \}$, respectively. That is, the overall validation $\mathcal{X}_{\textrm{II},n} = \mathcal{X}_{\textrm{II},n}^\textrm{Pilot} \cup \mathcal{X}_{\textrm{II},n}^\textrm{Adapt}$ corresponds to the optimal design, where $\mathcal{I} = \mathcal{I}^{\textrm{Pilot}} \cup \mathcal{I}^{\textrm{Adapt}}$. Similarly, we consider an adaptive constraint on the final validation size, $n = \sum_{(y,\delta,\mathbf{z})} \big[ n^{\textrm{Pilot}}(y,\delta,\mathbf{z}) + n^{\text{Adapt}}(y,\delta,\mathbf{z})\big]$, where $n^{\textrm{Pilot}}(y,\delta,\mathbf{z})$ and $n^{\textrm{Adapt}}(y,\delta,\mathbf{z})$ are sampling sizes on each $(y,\delta,\mathbf{z})$-stratum for the pilot and adaptive validation, respectively. We apply the Lagrangian multiplier method to minimize $\Sigma_{[k,k]}$ in Theorem \ref{asymp-normal} with respect to $\pi(Y, \Delta, \mathbf{Z})$ under the adaptive constraint. Then, the adaptive sampling design is given by
\begin{align}
	n^{\textrm{Adapt}}(y,\delta,\mathbf{z}) = \hat{n}^{\textrm{Opt}}(y,\delta,\mathbf{z}) - n^{\textrm{Pilot}}(y,\delta,\mathbf{z}) \label{adaptive-design},
\end{align} 
where $\hat{n}^{\textrm{Opt}}(y,\delta,\mathbf{z})$ is the estimated optimal sampling size of \eqref{optimal-design}. Here, the information matrix $I_V$ and the conditional variance of the score function ${\mathrm{Var}}\big(U_1(\boldsymbol{\theta}) | y, \delta, \mathbf{z} \big)$ can be consistently estimated using the individuals in the phase two pilot sample. We employ inverse probability weighting to estimate $I_V$ by
\begin{align}
	\widehat{I}_V = -\frac{1}{N}\sum_{i \in \mathcal{I}^{\textrm{Pilot}}} \frac{N(Y_i,\Delta_i,\mathbf{Z}_i)}{n^{\textrm{Pilot}}(Y_i,\Delta_i,\mathbf{Z}_i)} \frac{\partial^2}{\partial \boldsymbol{\theta} \partial \boldsymbol{\theta}^\top} \log L_1(\boldsymbol{\theta}; Y_i, \Delta_i, \mathbf{X}_i). \label{est-pilot1}
\end{align}
Also, $\mathrm{Var}\big(U_1(\boldsymbol{\theta}) | y, \delta, \mathbf{z} \big)$ is estimated by the sample covariance matrix  of the score function within each $(y,\delta,z)$-stratum such that
\begin{align}
    \widehat{\mathrm{Var}} \big(U_1(\boldsymbol{\theta}) | y, \delta, z \big)
        &= \frac{n^{\textrm{Pilot}}(y,\delta,\mathbf{z})}{n^{\textrm{Pilot}}(y,\delta,\mathbf{z})-1} \big\{ \hat{\mu}_2(\boldsymbol{\theta}; y,\delta,\mathbf{z}) - \hat{\mu}_1(\boldsymbol{\theta}; y,\delta,\mathbf{z})^2\big\}, \label{est-pilot2}
\end{align}
where $\hat{\mu}_\ell(\boldsymbol{\theta}; y,\delta,\mathbf{z}) = n^{\textrm{Pilot}}(y,\delta,\mathbf{z})^{-1} \sum_{i \in \mathcal{I}^{\textrm{Pilot}}} U_1(\boldsymbol{\theta} ; Y_i, \Delta_i, \mathbf{X}_i)^\ell \cdot \mathbb{I}(Y_i=y, \Delta_i=\delta, \mathbf{Z}_i = \mathbf{z})$, for $\ell=1,2$. Then, we evaluate \eqref{est-pilot1} and \eqref{est-pilot2} with the inverse probability weighted estimator of $\boldsymbol{\theta}$ based on the first phase sample $\mathcal{X}_{\textrm{I},N}$ and the pilot validation set $\mathcal{X}_{\textrm{II},n}^\textrm{Pilot}$.  The expressions for the score function and Hessian matrix of \eqref{log-linear} for two discussed choices of the survival models can be found in Supplementary Material Section \ref{technical-details}. Due to sparse observations or oversampled pilot data on some strata, we may not achieve practically the optimal design with \eqref{adaptive-design} when $N(y,\delta,\mathbf{z}) < \hat{n}^{\textrm{Opt}}(y,\delta,\mathbf{z})$ or $n^{\textrm{Pilot}}(y,\delta,\mathbf{z}) > \hat{n}^{\textrm{Opt}}(y,\delta,\mathbf{z})$. In either case, we set $n^{\textrm{Adapt}}(y,\delta,\mathbf{z}) = 0 \vee \big\{ N(y,\delta,\mathbf{z}) - n^{\textrm{Pilot}}(y,\delta,\mathbf{z}) \big\}$ for the saturated strata and distribute the remaining validation allocation to the other strata proportional to the estimated optimal sampling sizes. This approach for handling saturated strata is similar to that of \cite{mcisaac2015adaptive}, originally introduced by \cite{reilly1995mean}.

	\section{Numerical illustrations} \label{sec-numerical}

In this section, we examine the performance of our proposed mean score estimator and adaptive phase two sampling procedure first by a computer simulation study. We then further illustrate the method with an analysis of data from the National Wilms Tumor Study (NWTS). For this example, the original survival outcome was continuous and so in addition to presenting the discrete time analysis, we consider whether the proposed phase two sampling procedure provided efficiency gains for the analysis of the continuous time outcome. We also provide further discussion on phase two sampling in the setting of intermittently censored outcomes. Data and source code in R (version 3.6.1) for our numerical studies are provided at \url{https://github.com/kyungheehan/mean-score}.
	
	\subsection{Simulation Study} \label{subsec-sim}
Here, we evaluate the empirical performance of the proposed mean score estimator for the discrete survival time setting via a simulation study. We also evaluate the degree to which the proposed adaptive validation design improves estimation performance compared to simple random sampling and a balanced design for several scenarios. 	

We consider the conditional hazard model \eqref{phz-model} with the complementary log transformation $g_2(u) = -\log(1-u)$, where $\boldsymbol{\beta} = (\log(1.5), \log(0.7), \log(1.3), -\log(1.3))^\top$. We assume survival status is observed at discrete times $0< t_1 < t_2 < \cdots< t_{10}<\infty$. As previously mentioned in Section \ref{subsec-cox}, this model will estimate the same $\boldsymbol{\beta}$ as in the underlying continuous-time Cox proportional hazards model.  We first generate a four dimensional covariate vector $\mathbf{X} = (X_1, \ldots, X_4)^\top$, which consists of both continuous and binary variables. 

We simulate correlated covariates with a unit scale between $0$ and $1$, by first considering a multivariate normal random vector $\mathbf{W} = (W_1, \ldots, W_4)^\top$ with zero mean and $\textrm{Cov}(W_j,W_k) = 0.3^{|j-k|}$, so that we put $X_j = Q_j(\Phi(W_j))$ for $j=1,2$ and $X_j \sim \textrm{Bernoulli}(\Phi(W_j))$ for $j=3,4$, where $\Phi$ is the cumulative standard normal distribution function, and $Q_1$ and $Q_2$ are quantile functions of the beta distribution with pairs of the shape and rate parameters $(2,1.5)$ and $(3,3)$, respectively. By doing this, all $X_j$'s are correlated with each other, and particularly $X_1$ and $X_2$ are marginally beta random variables with $\textrm{Corr}(X_1, X_2) \approx 0.290$.  
Since continuous covariates are generally bounded in practice,  the simulated continuous covariates $(X_1, X_2)^\top$ represent standardized covariates over the range of observations into unit intervals. 

For discrete survival outcomes, we note that $P(T=t_j \,|\, \mathbf{X} = \mathbf{x}) = \lambda_j(\mathbf{x}) \prod_{k=1}^{j-1}(1-\lambda_k(\mathbf{x}))$ enables us to generate the discrete survival outcomes as multinomial random variables associated with covariates. To simplify the censoring mechanism in our simulation, we set a fixed censoring time $C=t_6$ as the maximum  follow-up for all subjects, so that we observe a truncated survival time $Y = \min\{T, t_6\}$ and $\Delta=\mathbb{I}(T \leq t_6)$, It is worth mentioning that the proposed method also allows a random censoring time $C$. An application of the method with a more complex censoring mechanism will be considered in Section \ref{subsec-data}. 

\begin{figure}[!t]
	\centering
	\includegraphics[width=0.329\textwidth]{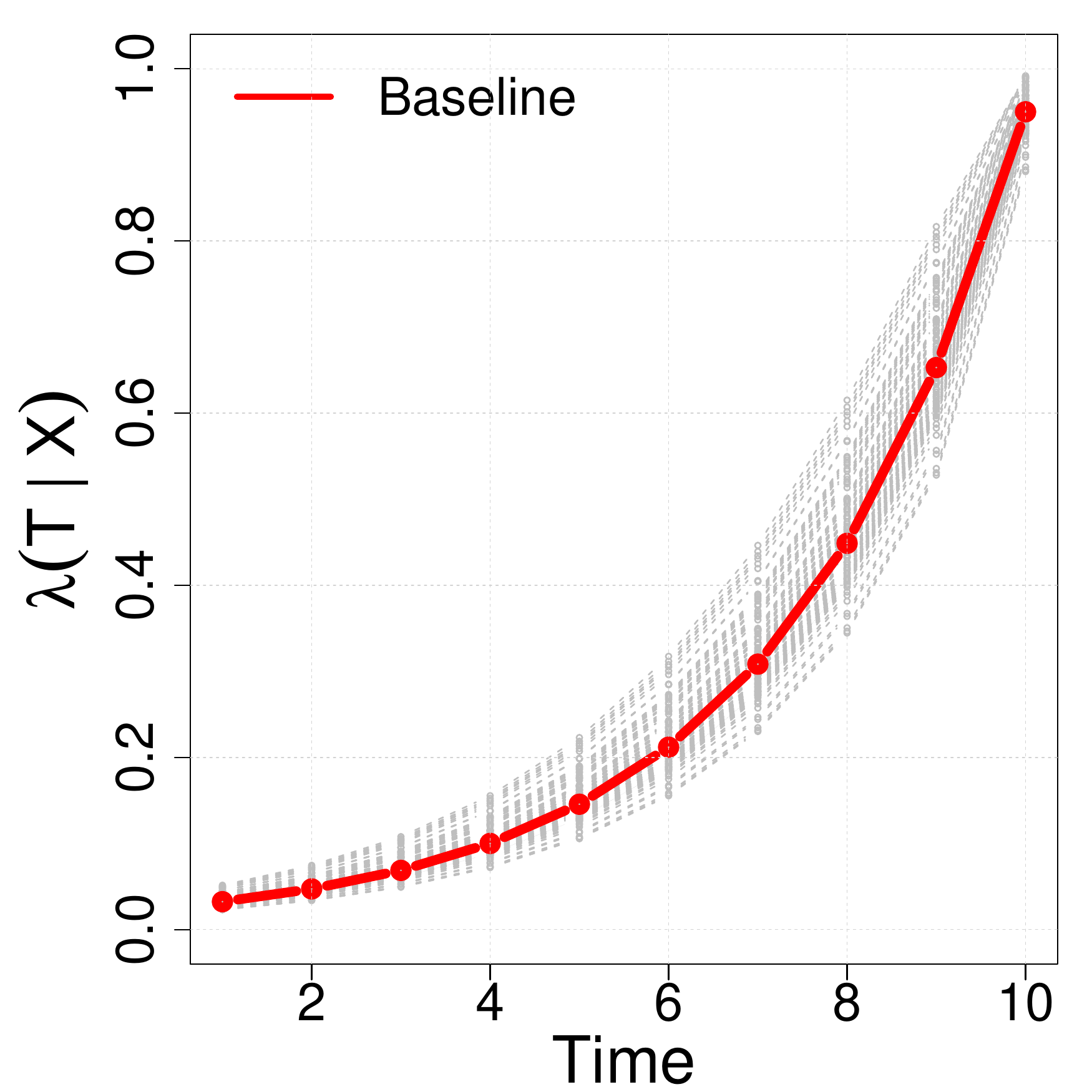}
	\includegraphics[width=0.329\textwidth]{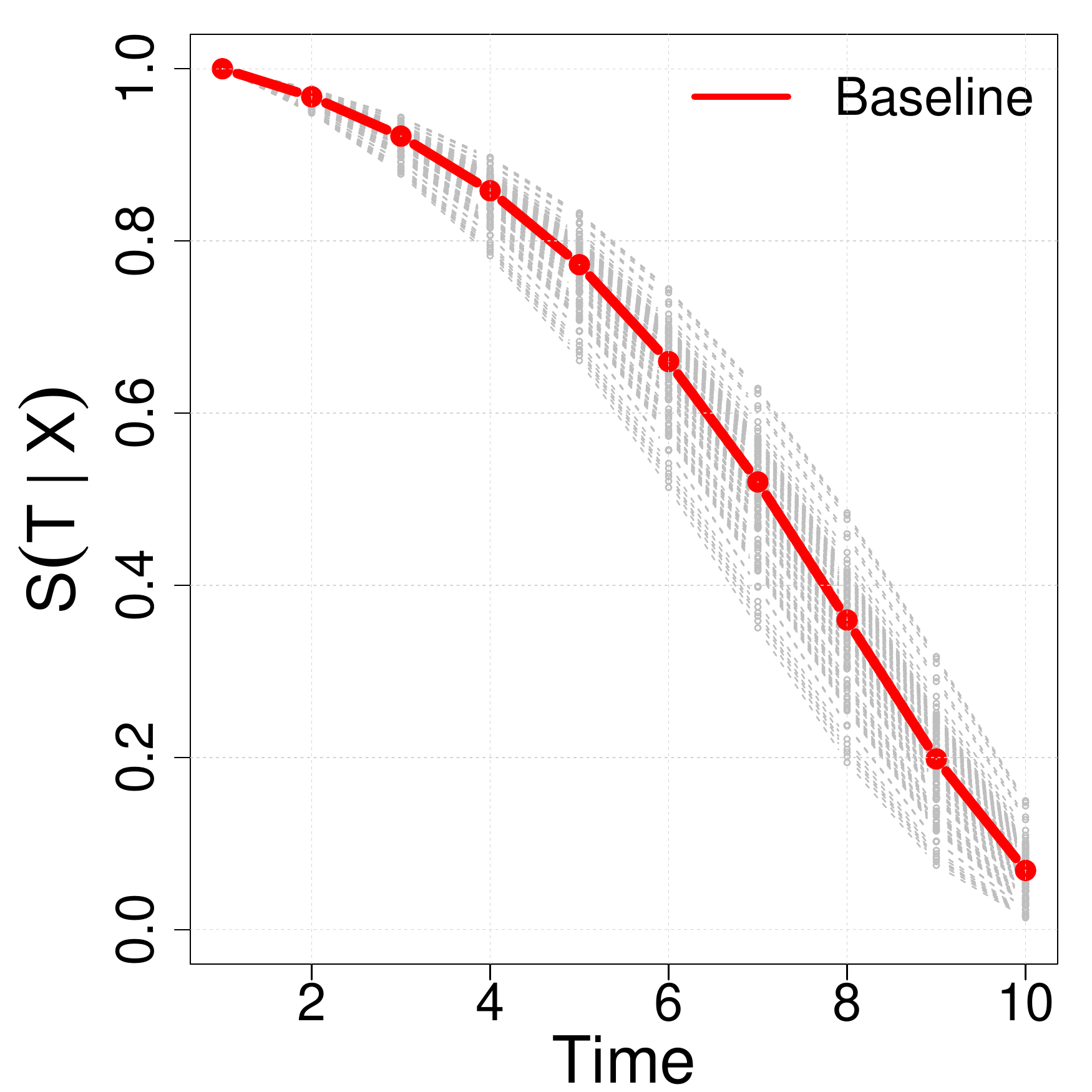}
	\includegraphics[width=0.329\textwidth]{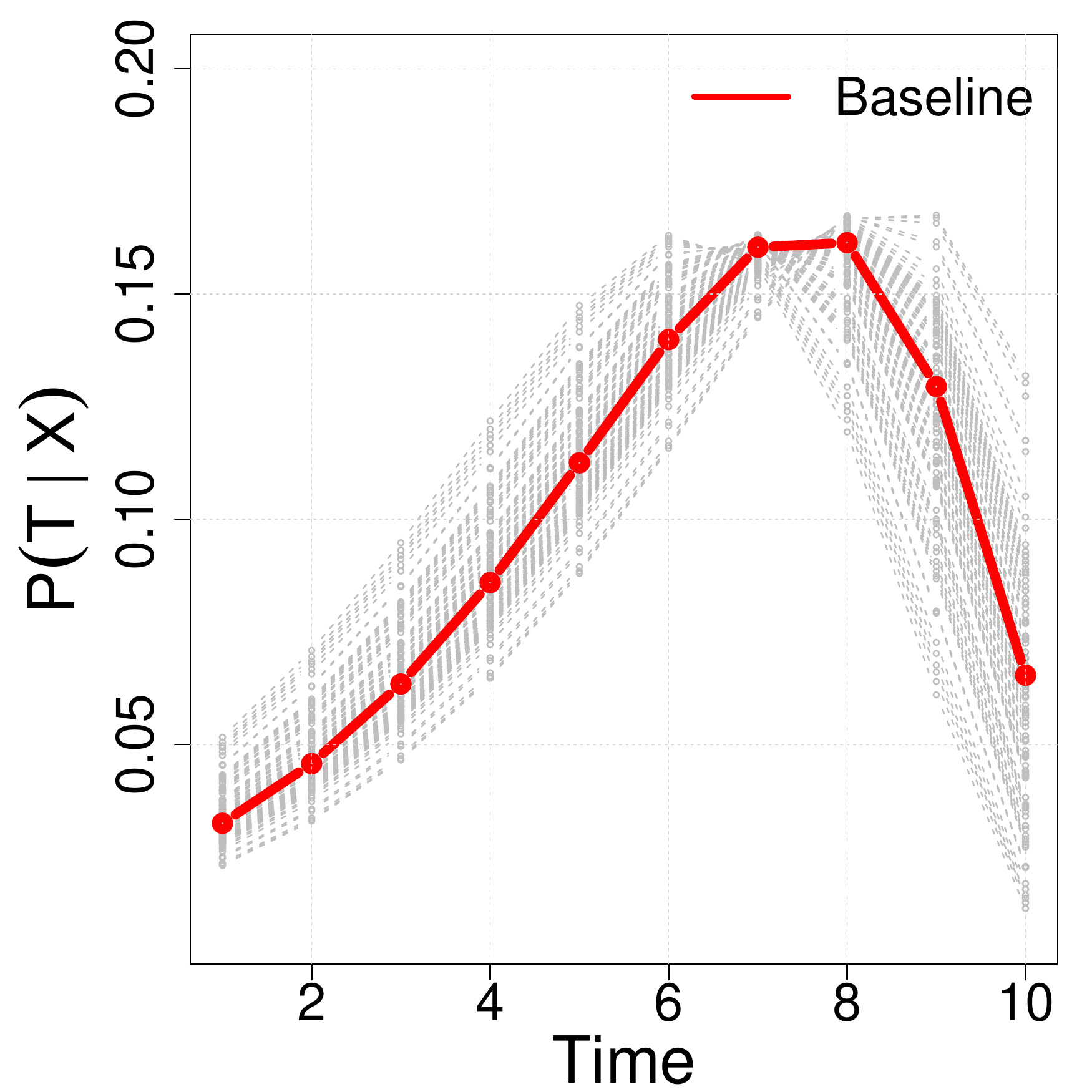}
	\caption{Illustrations of the simulation setting for the discrete-time survival model. The conditional hazard function (left) is demonstrated together with its associated survival and probability mass functions (middle and right), where $\lambda(t_j | \mathbf{x}) = P(T = t_j | T \geq t_j, \mathbf{X}=\mathbf{x})$, $S(t_j | \mathbf{x}) = P(T \geq t_j | \mathbf{X}=\mathbf{x})$ and $P(t_j | \mathbf{x}) = P(T=t_j | \mathbf{X}=\mathbf{x})$. The baseline functions are illustrated in red and gray solid lines show realization of conditional hazards, survival and probability functions associated with covariates for randomly chosen $100$ subjects. The marginal rate of censoring over the maximum follow-up period $t_6$ is $P(T > 6) \approx 0.5$.}
	\label{figure1}
\end{figure}

Next, we generate $N$-random copies $\mathcal{X}_{N} = \{ (Y_i, \Delta_i, \mathbf{X}_i) : 1 \leq i \leq N \}$ of $(Y,\Delta, \mathbf{X})$, which will inform the benchmark full cohort analysis. We consider full cohort sizes $N=2000, 4000, 8000$ in the simulation. By choosing different baseline hazards, we considered three scenarios of overall censoring rates with $30\%$, $50\%$ and $70\%$.  Here, we mainly refer to simulation results when the censoring rate is $50\%$, since we found that this setting fundamentally provides similar lessons with the other scenarios which can be found in the Supplementary Material Tables \ref{table-sim-cens30} and \ref{table-sim-cens70}. Figure \ref{figure1} illustrates the conditional hazard, survival and probability functions in our simulation settings. 

We assume that direct observations of $X_1$ are not available in the phase one sample of the cohort, but instead one observes a discretized and error-prone exposure $Z$ such that
\begin{eqnarray}
\begin{split}
	Z = 
		\left\{
		\begin{array}{cc}
			1	&	(0 < X_{1}^\ast \leq 0.25),\\
			2	&	(0.25 < X_{1}^\ast \leq 0.5),\\
			3	&	(0.5 < X_{1}^\ast \leq 0.75),\\
			4	&	(0.75 < X_{1}^\ast < 1),\\
		\end{array}
		\right.
\end{split} \label{sim-aux}
\end{eqnarray}
where $X_{1}^\ast = X_{1} + \varepsilon$ is perturbation of $X_{1}$ with an independent measurement error $\varepsilon \sim N(0,0.1^2)$. Let $Z^o$ be the true discretization of $X_{1}$ defined similarly to $Z$ by replacing $X_{1}^\ast$ with $X_{1}$. For the assumed parameter values, the discordance or misclassified rate between $Z^o$ and $Z$ was $P(Z^o \neq Z) \approx 0.284$ and this shows that $Z$ is not only a discrete but also an error-prone surrogate of $X_1$, and potentially associated with $(X_2,X_3,X_4)$. Finally, we set $\mathcal{X}_{\textrm{I},N} = \{ (Y_i, \Delta_i, Z_i) :  1 \leq i \leq N \}$ to be the phase one sample available on all subjects. 

Efficient estimation of the regression coefficient of $X_1$ is of interest and the optimal sampling allocations are designed to minimize the variance of the $\beta_1$ estimate in our simulation study. Since the optimal sampling design \eqref{optimal-design} depends on nuisance parameters defined by the population structure, namely $I_V$ and $\mathrm{Var}\big(U_1(\boldsymbol{\theta}) | y, \delta, z \big)$ in Theorem \ref{asymp-normal}, we approximate their true values with empirical estimates obtained from an externally generated large sample of size $N_0=10^4$ that was independent from the full cohort $\mathcal{X}_N$. We define the optimal sampling strategy based on these values as the oracle procedure and refer to Supplementary Material Section \ref{technical-details} for some technical details for the derived optimal allocation. In practice, however, the oracle procedure is infeasible and so we are interested in evaluating the adaptive sampling design as described in Section  \ref{sec-optsamp}. For this, we first sample a pilot validation subset $\mathcal{X}_{\textrm{II},n}^{\textrm{Pilot}}$ and estimate the optimal sampling design together with the phase one sample $\mathcal{X}_{\textrm{I},N}$. To accommodate all possible strata information in the first phase sample, we employ balanced sampling for the pilot study, with $n^{\textrm{Pilot}}(Y_i,\Delta_i,Z_i)$ equal for all $\{Y_i,\Delta_i,Z_i\}$. We then estimate $I_V$ and  ${ \mathrm{Var}}\big(U_1(\boldsymbol{\theta}) | y, \delta, z \big)$, as outlined at the end of the previous section. Next, we draw an additional validation subset $\mathcal{X}_{\textrm{II},n}^{\textrm{Adapt}}$ for each stratum following \eqref{adaptive-design}, and the mean score method is finally applied to the two-stage analysis for the phase one sample $\mathcal{X}_{\textrm{I},N}$ and the adaptive phase two sample $\mathcal{X}_{\textrm{II},n} = \mathcal{X}_{\textrm{II},n}^{\textrm{Pilot}} \cup \mathcal{X}_{\textrm{II},n}^{\textrm{Adapt}}$. We considered validation subset sizes of $n=200, 400, 800$ and took equal proportions for the pilot and adaptive samples. 

Due to saturated strata, there might be some remaining validation size to be allocated, for example $n/2-\sum_{(y,\delta,z)} n^{\textrm{Pilot}}(y,\delta,z) > 0$ in the pilot study. In this case, we randomly select unvalidated subjects for the remaining allocation, which is also similarly applied to the adaptive sample. We note that the proportion of the pilot sample size to the adaptive entails a trade-off between precision of the nuisance parameters needed for optimal sampling design and efficiency gains from the adaptive validation when the final phase two sample size is fixed. Some preliminary simulations showed that the adaptive sampling design with nearly equal sizes of the pilot and adaptive samples usually produced robust and efficient estimates (data not shown), which is similar to observations made by \cite{mcisaac2015adaptive} for the two-stage analysis with binary outcomes. 

We compare our proposed adaptive sampling method, which we refer to here as mean score adaptive (MS-A), to the mean score method using the oracle design (MS-O) and to some other standard estimation methods for two-phase designs. The complete case analysis of fully randomly selected $n$-validation sample (CC-SRS) will give unbiased results in likelihood-based inference. Since CC-SRS does not use auxiliary information of the first stage sample, we  examine if the optimal design for the mean score method improves estimation performance of CC-SRS and evaluate efficiency gains from the two-stage analysis (MS-SRS). We also conduct design-based estimation with balanced sampling such that validation size is equally allocated to each $(y,\delta,z)$-stratum of the first stage sample. Similarly to the proposed adaptive procedure, if there are some remaining individuals to be allocated after balanced sampling, due to saturated strata, we randomly sample the remaining from the unvalidated subjects to yield a final total phase two sample of $n$ individuals. The design-based estimation with balanced sampling is given by a Horvitz-Thompson type estimator (MS-BAL), where sampling proportions of validation within strata are used for inverse probability weights as in \eqref{em-re}. For our setting, we note that the inverse probability weighted (IPW) estimator is technically the same with the design-based mean score estimator which incorporates the balanced sampling weights pre-specified in the two-phase analysis \citep{mcisaac2014response}.  We implement the proposed mean score estimator with adaptive sampling described above, estimating the necessary nuisance parameters from the validation subset (MS-A) and the when using the oracle design (MS-O). 
Finally, we consider the full cohort analysis (Full-CC) based on fully observed data, as the benchmark performance, which empirically gives the upper bound of efficiency for the two-phase analysis, since the Full-CC uses the complete covariate information on all subjects.

\begin{table}[!t]
    \begin{center}
    \caption{Relative performance for the estimation of $\beta_1$ is compared for (i) the complete case analysis with simple random sampling (CC-SRS), (ii) the mean score method with simple random sampling (MS-SRS), (iii) a design-based estimation with balanced sample, equivalent to the mean score (MS-BAL), (iv \& v) the mean score estimation with adaptive sampling (MS-A) and the optimal sampling design (MS-O), for varying sample sizes. Results for the full cohort estimator based on complete data are provided as a benchmark.  Mean squared error (MSE) and its bias-variance decomposition are estimated from $1000$ Monte Carlo replications, where the censoring rate was $50\%$. The adaptive and optimal sampling designs were for efficient estimation of $X_1$ with $\beta_1=\log(1.5) \approx 0.405$. In all adaptive sampling scenarios, we took equal proportions for the pilot and adaptive samples.}
    \label{table-sim-cens50}
    \small
    \begin{tabular}{ccc  rrr c rrr }
	\hline
	\multicolumn{1}{c}{\multirow{3}{*}{Sampling}}	& 	\multicolumn{1}{c}{\multirow{3}{*}{Estimation}}	&	\multirow{3}{*}{Criterion}	&	\multicolumn{7}{c}{Estimation performance by sample sizes}\\
	\cline{4-10}
		&	&	&	\multicolumn{3}{c}{\scriptsize{$N=4000$}}	& &	\multicolumn{3}{c}{\footnotesize{$n=400$}}\\
	\cline{4-6} \cline{8-10}
		&	&	&	\footnotesize{$n=200$}	&	\footnotesize{$n=400$}	&	\footnotesize{$n=800$}	& &	\scriptsize{$N=2000$}	&	\scriptsize{$N=4000$}	&	\scriptsize{$N=8000$}\\
	\hline
	\multirow{3}{*}{Full cohort}		
			&	\multirow{3}{*}{CC}
						&	$\sqrt{\textrm{MSE}}$	&	0.094	&	0.094	&	0.094	& &		0.129 	&	0.094	&	0.067 	\\
		\multicolumn{2}{c}{}	&	Bias					&	0.003	&	0.003	&	0.003	& &		0.004	&	0.003	&	0.000 	\\
		\multicolumn{2}{c}{}	&	$\sqrt{\textrm{Var}}$		&	0.094	&	0.094	&	0.094	& &		0.129 	&	0.094	&	0.067	\\
	\hline
	 \multirow{6}{*}{SRS}		
			&	\multirow{3}{*}{CC}
						&	$\sqrt{\textrm{MSE}}$	&	0.470	&	0.330	&	0.228	& &		0.324 	&	0.330	&	0.321 	\\
		\multicolumn{2}{c}{}	&	Bias					&	0.017	&	0.014	&	0.005	& &		0.011	&	0.014	&	-0.010 	\\
		\multicolumn{2}{c}{}	&	$\sqrt{\textrm{Var}}$		&	0.470	&	0.329	&	0.228	& &		0.324 	&	0.329	&	0.321 	\\
	\cline{2-10}
		&	\multirow{3}{*}{MS}
						&	$\sqrt{\textrm{MSE}}$	&	0.332	&	0.220	&	0.155	& &		0.237 	&	0.220	&	0.200 	\\
		\multicolumn{2}{c}{}	&	Bias					&	0.053	&	0.035	&	0.004	& &		0.020 	&	0.035	&	0.027 	\\
		\multicolumn{2}{c}{}	&	$\sqrt{\textrm{Var}}$		&	0.330	&	0.217	&	0.155	& &		0.236 	&	0.217	&	0.198 	\\
	\hline
	\multirow{3}{*}{Balanced}		
			&	\multirow{3}{*}{MS}
						&	$\sqrt{\textrm{MSE}}$	&	0.400	&	0.278	&	0.194	& &		0.276 	&	0.278	&	0.265 	\\
		\multicolumn{2}{c}{}	&	Bias					&	0.042	&	0.024	&	0.010	& &		0.026 	&	0.024	&	0.021 	\\
		\multicolumn{2}{c}{}	&	$\sqrt{\textrm{Var}}$		&	0.398	&	0.277	&	0.194	& &		0.275 	&	0.277	&	0.264 	\\
	\hline
	 \multirow{3}{*}{Adaptive}		
			&	\multirow{3}{*}{MS}
						&	$\sqrt{\textrm{MSE}}$	&	0.374	&	0.197	&	0.147	& &		0.214 	&	0.197	&	0.190 	\\
		\multicolumn{2}{c}{}	&	Bias					&	0.049	&	0.007	&	0.006	& &		0.004 	&	0.007	&	0.002 	\\
		\multicolumn{2}{c}{}	&	$\sqrt{\textrm{Var}}$		&	0.371	&	0.197	&	0.147	& &		0.214 	&	0.197	&	0.189 	\\
	\hline
	\multirow{3}{*}{Oracle}		
			&	\multirow{3}{*}{MS}
						&	$\sqrt{\textrm{MSE}}$	&	0.253	&	0.182	&	0.133	& &		0.202 	&	0.182	&	0.174 	\\
		\multicolumn{2}{c}{}	&	Bias					&	0.009	&	0.003	&	0.005	& &		0.012	&	0.003	&	-0.005 	\\
		\multicolumn{2}{c}{}	&	$\sqrt{\textrm{Var}}$		&	0.252	&	0.182	&	0.133	& &		0.202 	&	0.182	&	0.174 	\\
	\hline 
	\end{tabular}
    \end{center}
\end{table}

We investigated two different aspects of varying sample sizes and conducted six simulation scenarios; (i) increasing the phase two sample size from $n=200, 400, 800$ with a fixed full cohort size $N=4000$, (ii) increasing the full cohort size from $N=2000, 4000, 8000$ when the phase two sample size is fixed by $n=200$. Table \ref{table-sim-cens50} summarizes the relative performance for estimation of $\beta_1$. Compared to CC-SRS, MS-SRS had a reduction of the variance, which demonstrates efficiency gains of the mean score method from employing auxiliary information of the phase one sample. For $n=400$ and 800, MS-A was more efficient than MS-SRS, whereas for $n=200$, MS-SRS was more efficient. This suggests that $n=200$ is too small in this setting to gain efficiency using adaptive sampling; this is reasonable as the optimal sampling allocation is based on estimates derived from a pilot sample half that size. MS-BAL had relatively inferior performance compared to MS-SRS and MS-A, and this was true across all scenarios studied. The MS-O had the smallest mean squared error (MSE) in all scenarios, as expected, and it turned out that the superior performance of MS-O mostly came from variance reduction. As the cohort sample size increased, the associated adaptive sampling design (MS-A) behaved closely to MS-O, which indicates that the proposed adaptive approach  consistently approximates the oracle procedure. In the scenarios studied, increasing the phase two sample size was much more beneficial for efficiency gains for MS-A and MS-O, compared to the sample size increment of the phase one study. For example, MS-A achieved $60.7\% (\approx 1-0.147/0.374)$ variance reduction while the validation rate increased four-fold from $n=200$ to $n=800$ given $N=4000$ in the left three columns of Table \ref{table-sim-cens50}. On the contrary, increasing the phase one sample size from $N=2000$ to $N=8000$ only produced $11.2\% (\approx 1-0.190/0.214)$ improvement in the right three columns when $n=400$ is fixed. This suggests that the performance of the proposed MS-A  procedure is sensitive to the size of the phase two study, more so than the total cohort size for a fixed $n$. 

Simulation results for all regression parameters and various combinations of $n=200,400,800$ and $N=2000, 4000, 8000$ can be found in Supplementary Material Tables \ref{table-sim-app-n200-N4000}--\ref{table-sim-app-n400-N8000}. Comparing the relative performance of the methods for the other  regression parameters, including those for the discrete baseline hazards, we found that the MS-O outperformed the other estimators for all model parameters.  The oracle sampling design is infeasible in most practical situations; however, the two mean score estimators consistently showed the best performance for the target parameter among all other practical competitors. MS-A generally achieved the minimum mean squared errors for all parameters when the phase two sample size exceeded 200, compared to the other practical estimators, for the scenarios studied (Supplementary Material Tables \ref{table-sim-app-n200-N4000}--\ref{table-sim-app-n400-N8000}). For the setting where $n=200$, the MS-SRS performed the best amongst the practical estimators. These results suggest for that MS-A will perform well with respect to MSE for all model parameters, particularly the target and discrete proportional hazard parameters, for robust phase two sample size. For small phase two samples and similar censoring rates, the MS-SRS may be preferred.

	\subsection{Data example: The National Wilms Tumor Study} \label{subsec-data}
	
 Wilms tumor is a rare renal cancer occurring in children, where tumor histology and the disease stage at diagnosis  are two important risk factors for relapse and death. We consider data, reported by \cite{kulich2004improving}, on 3915 subjects  from two randomized clinical trials from the National Wilms Tumor Study (NWTS) \citep{d1989treatment,green1998comparison}. There are two measures of tumor histology, which is classified as either favorable (FH) or unfavorable (UH), one by a local pathologist and the other by an expert pathologist from a central facility. Because of the rarity of disease, local pathologists may be less familiar with Wilms Tumor and their assessment is subject to misclassification. The central assessment is considered to be the gold standard, or true histology in our analysis, and the local evaluations are considered surrogate observations for the central evaluation. Since histology for all subjects was validated by the central laboratory, NWTS data has been widely used to evaluate two-phase sampling methodology, such as by \cite{breslow1999design}, \cite{kulich2004improving} and \cite{lumley2011complex}. 

We demonstrate our proposed mean score method for discrete-time survival in an analysis of relapse in this NWTS cohort. We assume that the local histology is available for all subjects in a first phase sample and that only a sub-cohort was sampled for evaluation by the central pathologist in a second phase sample. Specifically, we are interested in the proportional hazards discrete-time survival model \eqref{phz-model} for time to relapse, under the complementary log transformation $g_2(u) = -\log(1-u)$, in order to evaluate the risk associated with unfavorable central histology, late (III/IV) disease stage versus the early (I/II) stage, age at diagnosis (year), and tumor  diameter (cm). For this model, we also include an interaction between histology and stage of disease. 

In this cohort, $90\%$ of the $669$ events occurred within the first three years of diagnosis, while less than $5\%$ of non-relapsed subjects were censored in the same period. Based on this observation, we first define a modified, or reduced, cohort to include only patients who had an event or were fully followed up in the first 3 years, so that censoring only occurs at the third year ($82.2\%$), the assumed end of the study. This modified cohort included $N=3757$ subjects and is used in our NWTS data analysis. We consider the regression coefficients from the  discrete-time survival analysis of the modified full cohort as the reference values. We took this approach first out of concern that the large number of nuisance parameters introduced by the small number of individuals with intermittent censored outcomes, due to the added strata for a binary outcome at each failure time interval, might adversely affect the mean score method. We will evaluate efficiency gains of the optimal mean score design, based on the discretized survival outcomes, for the continuous-time analysis in Section \ref{design-cox}. Finally, we also conducted an analysis of all individuals, regardless of the time of censoring, and discuss design issues regarding how to handle intermittent censoring in our framework in Section \ref{design-censoring}.

 We first discretize the continuous event time into six 6-month intervals, so that we model the hazard of relapse during the first three years after diagnosis. As in Section \ref{subsec-sim}, we consider four different sampling scenarios for the phase two subsample: simple random sampling, balanced allocation across strata, the adaptive and optimal (oracle) mean score designs, where the last three employ stratified sampling based on the phase one sample. To evaluate the efficiency gains of the proposed mean score approach,  we performed the two-stage analysis, with a phase two sample of $n=400$, 1000 times. For implementation of the optimal design, we estimate parameters in \eqref{optimal-design} using the oracle procedure by using the central histology records of the full cohort. We refer to Section \ref{subsec-sim} for further details of implementation. Bias and efficiency are calculated using the estimates of the full cohort analysis with the central histology as the reference.

Table \ref{table-wilms-n400} demonstrates the performance of the different methods, where efficient estimation of the interaction between unfavorable histology and late stage of the disease is of main interest. MS-A outperformed the other practical competitors: CC-SRS, MS-SRS and MS-BAL. For example, MS-A had a $38.7\% (\approx 1-0.425/0.693)$ variance reduction compared to CC-SRS for estimating the interaction term. We also note that performance of MS-A was pretty close to the oracle procedure, MS-O; MS-A was only $7\%$ less efficient than MS-O for estimating the regression coefficient of interest. Furthermore, across all parameters, MS-A achieved the smallest mean squared error among the practical competitors, not only for baseline hazards estimation but also regression coefficients. Overall, the proposed method performed better than the other two-phase estimators, with both an efficient design and by incorporating auxiliary information from the phase one sample.

\begin{table}[!t]
    \begin{center}
    \caption{For the discrete-time survival analysis of the National Wilms Tumor Study, we compare five different methods: (i) Complete case analysis with simple random sampling (CC-SRS), (ii) mean score estimation with simple random sampling (MS-SRS), (iii) mean score estimation with a balanced sample (MS-BAL), also equivalent to the Horvitz-Thompson estimator, (iv) mean score estimation with adaptive sampling (MS-A) and (v) mean score estimation with the optimal sampling design (MS-O). The optimal sampling design was estimated using the full cohort data. The MS-A and MS-O designs are for efficient estimation of the interaction effect between unfavorable histology and disease stage.  Results from the full cohort analysis with complete data are presented as a benchmark. Mean squared error and its bias-variance decomposition are estimated using $1000$ phase two samples of $n=400$ from the reduced full cohort ($N=3757$). We took equal proportions for the pilot and adaptive samples.}
    \label{table-wilms-n400}
    \small
    \resizebox{\columnwidth}{!}{%
    \begin{tabular}{ccc rrrrrr c rrrrr}
	\hline
	\multicolumn{1}{c}{\multirow{2}{*}{Sampling}}	& 	\multicolumn{1}{c}{\multirow{2}{*}{Estimation}}	&	\multirow{2}{*}{Criterion}	&	\multicolumn{6}{c}{Baseline hazard in complementary log-log scale}		&	&	\multicolumn{5}{c}{Regression coefficient}\\
	\cline{4-9}\cline{11-15}
		&	&	&	0.5yr	&	1yr	&	1.5yr	&	2yr	&	2.5yr	&	3yr	&	&	UH$^1$	&	Stage$^2$	&	Age$^3$	&	dTmr$^4$	&	U$\ast$S$^5$\\
	\hline	
	\multicolumn{1}{c}{Full cohort} 	&	CC	&	Estimate	&	-4.028	&	-3.876	&	-4.336	&	-5.005	&	-5.353	&	-5.719	&	&	1.058	&	0.280	&	0.063	&	0.032	&	0.636\\
	\hline
	 \multirow{6}{*}{SRS}		
			&	\multirow{3}{*}{CC}
						&	$\sqrt{\textrm{MSE}}$	&	0.452 & 0.456 & 0.458 & 0.666 & 1.459 & 2.468	&	&	0.555 & 0.317 & 0.046 & 0.031 & 0.693 \\
		\multicolumn{2}{c}{}	&	Bias					&	-0.071 & -0.065 & -0.077 & -0.096 & -0.278 & -0.649	&	&	-0.015 & 0.018 & -0.002 & 0.001 & 0.032 \\
		\multicolumn{2}{c}{}	&	$\sqrt{\textrm{Var}}$		&	0.446 & 0.451 & 0.452 & 0.659 & 1.432 & 2.381	&	&	0.555 & 0.317 & 0.046 & 0.031 & 0.692 \\
	\cline{2-15}
		&	\multirow{3}{*}{MS}
						&	$\sqrt{\textrm{MSE}}$	&	0.420 & 0.414 & 0.411 & 0.558 & 1.375 & 2.441	&	&	0.548 & 0.337 & 0.047 & 0.034 & 0.732 \\
		\multicolumn{2}{c}{}	&	Bias					&	-0.041 & -0.034 & -0.042 & -0.101 & -0.291 & -0.757	&	&	-0.091 & 0.017 & -0.001 & 0.002 & 0.094 \\
		\multicolumn{2}{c}{}	&	$\sqrt{\textrm{Var}}$		&	0.418 & 0.412 & 0.409 & 0.549 & 1.344 & 2.320	&	&	0.541 & 0.337 & 0.047 & 0.034 & 0.726 \\
	\hline
	\multirow{3}{*}{Balanced}		
			&	\multirow{3}{*}{MS}
						&	$\sqrt{\textrm{MSE}}$	&	0.410 & 0.404 & 0.396 & 0.391 & 0.389 & 0.388	&	&	0.370 & 0.353 & 0.056 & 0.035 & 0.554 \\
		\multicolumn{2}{c}{}	&	Bias					&	-0.109 & -0.096 & -0.084 & -0.076 & -0.072 & -0.070	&	&	0.043 & 0.007 & 0.013 & 0.005 & 0.004 \\
		\multicolumn{2}{c}{}	&	$\sqrt{\textrm{Var}}$		&	0.396 & 0.393 & 0.387 & 0.383 & 0.382 & 0.382	&	&	0.368 & 0.353 & 0.054 & 0.035 & 0.554 \\
	\hline
	 \multirow{3}{*}{Adaptive}		
			&	\multirow{3}{*}{MS}
						&	$\sqrt{\textrm{MSE}}$	&	0.315 & 0.309 & 0.303 & 0.299 & 0.297 & 0.296	&	&	0.284 & 0.254 & 0.042 & 0.025 & 0.425 \\
		\multicolumn{2}{c}{}	&	Bias					&	-0.065 & -0.058 & -0.052 & -0.048 & -0.046 & -0.045	&	&	0.003 & -0.007 & 0.008 & 0.003 & 0.034 \\
		\multicolumn{2}{c}{}	&	$\sqrt{\textrm{Var}}$		&	0.308 & 0.304 & 0.298 & 0.295 & 0.294 & 0.293	&	&	0.284 & 0.254 & 0.041 & 0.025 & 0.424 \\
	\hline
	 \multirow{3}{*}{Oracle}		
			&	\multirow{3}{*}{MS}
						&	$\sqrt{\textrm{MSE}}$	&	0.311 & 0.306 & 0.301 & 0.296 & 0.299 & 0.295	&	&	0.250 & 0.256 & 0.038 & 0.025 & 0.396 \\
		\multicolumn{2}{c}{}	&	Bias					&	-0.048 & -0.043 & -0.038 & -0.035 & -0.042 & -0.036	&	&	0.010 & 0.003 & 0.004 & 0.001 & 0.013 \\
		\multicolumn{2}{c}{}	&	$\sqrt{\textrm{Var}}$		&	0.307 & 0.303 & 0.298 & 0.294 & 0.296 & 0.293	&	&	0.250 & 0.256 & 0.037 & 0.025 & 0.396 \\
	\hline 
	\end{tabular}
	}
    \end{center}
    \vspace{-0.75cm}
    \footnotesize  
    \begin{flushleft}
    	$^1$ Unfavorable histology versus favorable; $^2$ disease stage III/IV versus I/II;\\
	$^3$ year at diagnosis; $^4$ tumor diameter (cm); $^5$ interaction effect between UH and Stage.
    \end{flushleft}
\end{table}

\begin{figure}[!t]
	\centering
	\begin{tabular}{cc}
		(a) Unfavorable histology (UH) \hspace{1cm}	&	\hspace{-1.5cm} (b) Late stage of disease \\
		\includegraphics[width=0.45\textwidth]{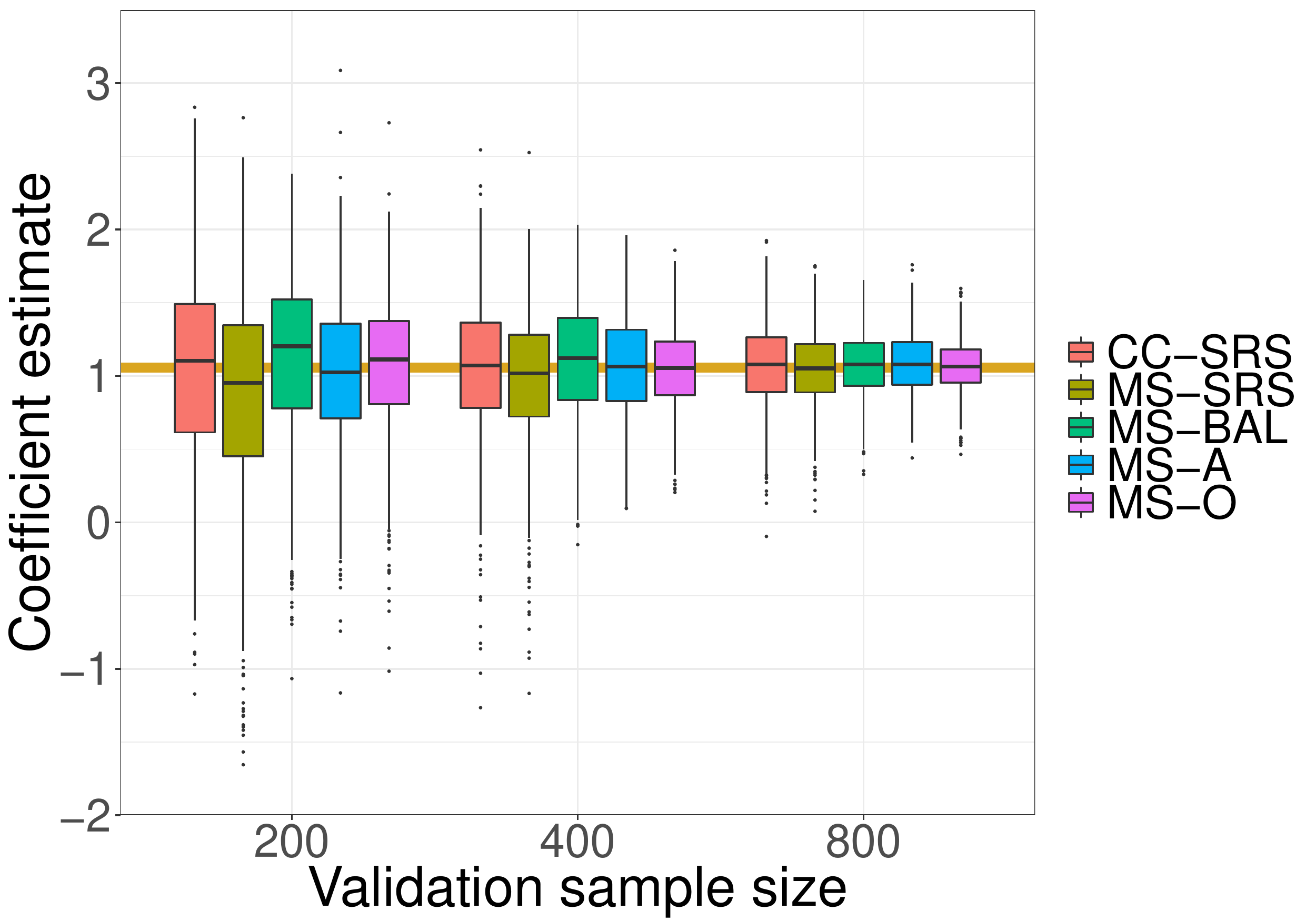} & \includegraphics[width=0.45\textwidth]{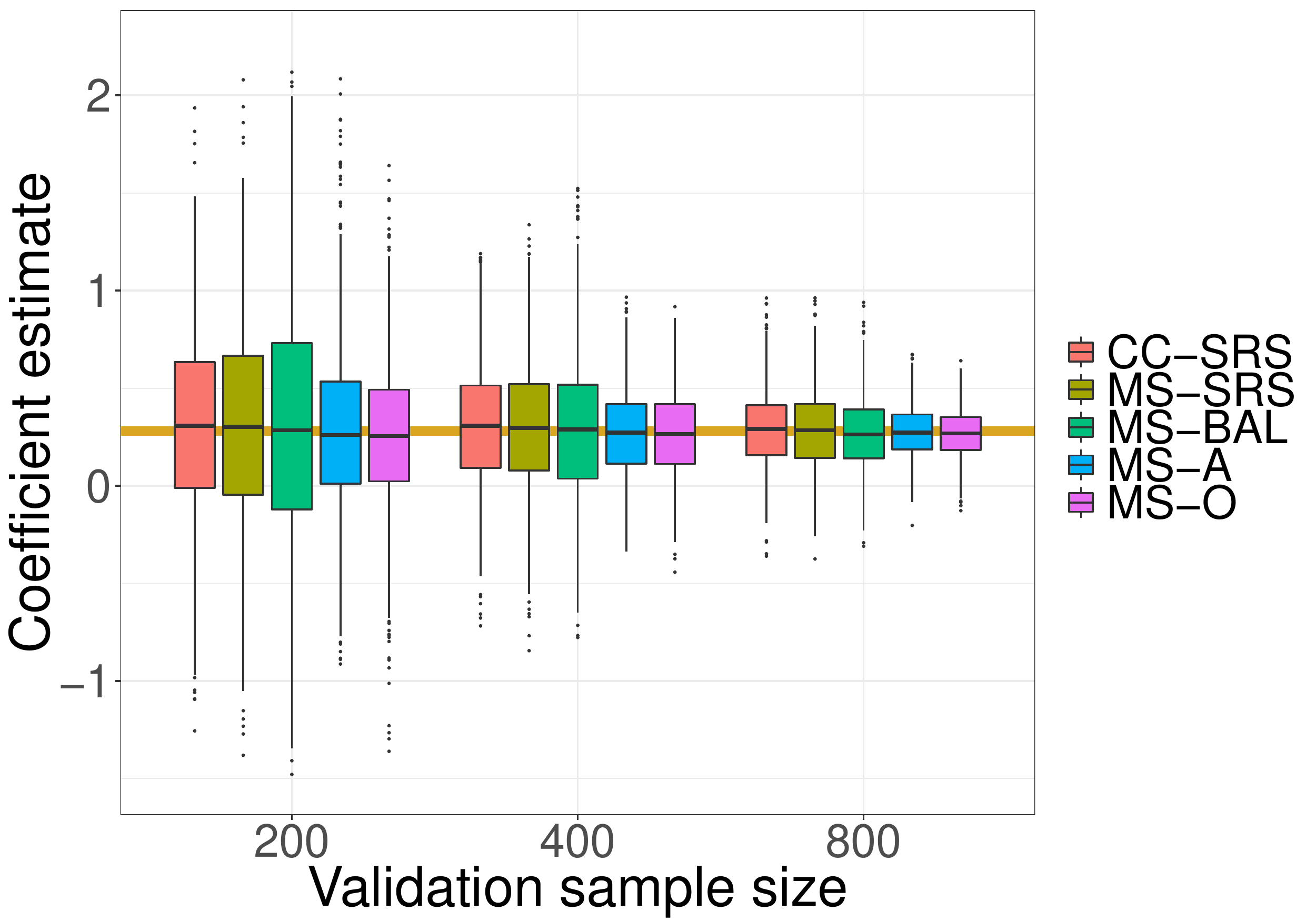}\\\\
		(c) Age \hspace{1cm}	&	\hspace{-1.25cm} (d) Tumor diameter \\
		\includegraphics[width=0.45\textwidth]{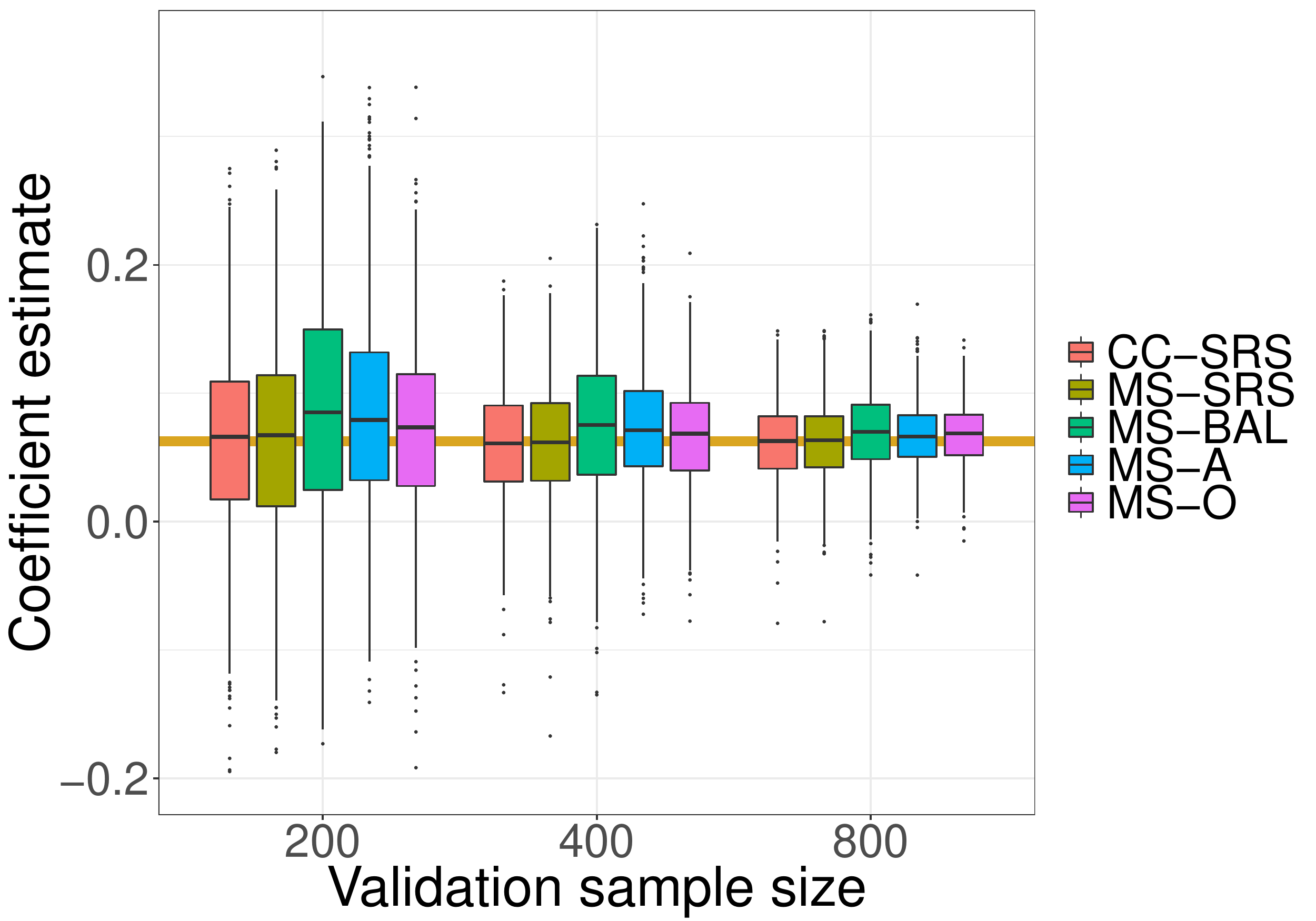} & \includegraphics[width=0.45\textwidth]{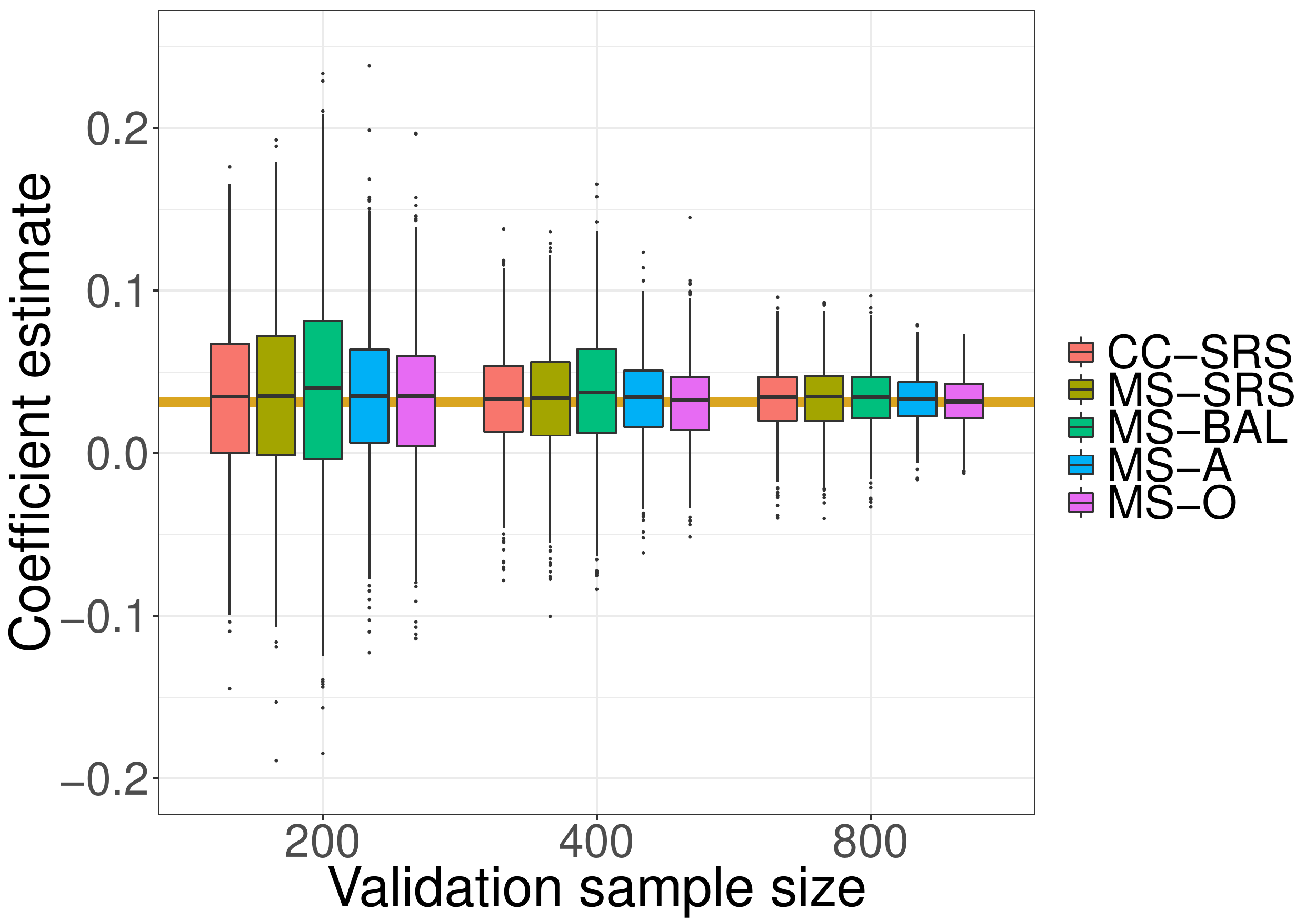}\\\\
		\multicolumn{2}{c}{\hspace{-1.25cm}(e) UH$\times$(Late stage of disease)}\\
		\multicolumn{2}{c}{\includegraphics[width=0.45\textwidth]{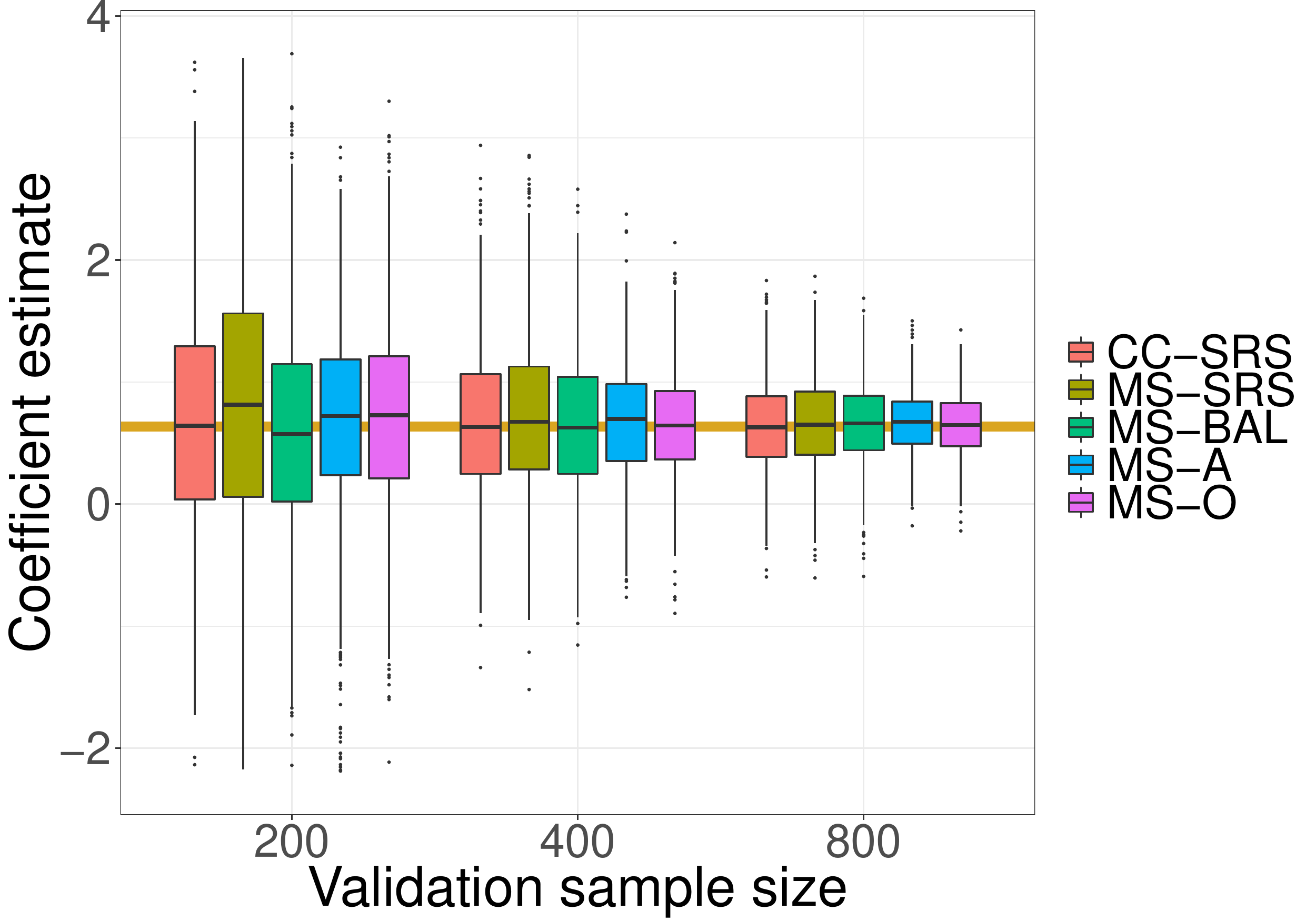}}
\end{tabular}
\caption{We demonstrate the relative performance of different methods by repeatedly subsampling the phase two sample $1000$ times from the reduced full cohort ($N=3757$) of the National Wilms Tumor Study data, while varying the target parameter for efficient estimation and the phase two sample size. Efficient results are shown for (a) unfavorable histology, (b) late stage of the disease, (c) age of diagnosis (year), (d) tumor diameter (cm), and the interaction between the histology and late stage, respectively, when validation sizes are $n=200, 400, 800$. The yellow horizontal lines represent the reference parameter estimates obtained from the full cohort analysis.}
\label{figure2}
\end{figure}

In Figure \ref{figure2}, we examine the relative performance of the different estimators for each of the regression parameters and phase two sample sizes $n=200, 400, 800$. In the top-left panel (a), we show the estimation results for the regression coefficient for unfavorable histology (UH) over 1000 repetitions of subsampling, when the adaptive and optimal mean score allocations were designed for the efficient estimation of UH. The rest of the panels are similar, showing results for the other regression coefficients they were set as targets, namely when the MS-A and MS-O designs were for efficient estimation of (b) late stage of the disease, (c) age of diagnosis (year), (d) tumor diameter (cm) and (e) the interaction between UH and stage of the disease, respectively. MS-O consistently showed superior performance for all phase two ample sizes; however, MS-A again had the smallest mean squared error among the practical methods. Further, the performance of MS-A tended to be close to MS-O as the validation size increased. 

	\subsection{Continuous-time survival analysis with the mean score design} \label{design-cox}

In many practical settings, the continuous-time Cox model will be the analysis of primary interest. We further investigate benefits of the proposed mean score sampling design, when the phase two estimating equation employs the continuous-time Cox proportional hazards model.  In Section \ref{subsec-cox}, we discussed the direct connection between the the continuous-time Cox model \eqref{phz-model} and the analogous grouped discrete-time model with the complementary log transformation, in the sense that the two models have the same regression coefficient $\boldsymbol{\beta}$.  We can conduct the two-phase analysis of the Cox model with the proposed optimal mean score sampling method derived for the parameter of interest in the discretized model.

\begin{table}[!b]
    \begin{center}
    \caption{Performance the two-phase continuous-time Cox model analysis of time to relapse in the National Wilms Tumor Study. We used inverse probability weights (IPW) for the two-phase analysis for four different sampling designs for the second phase; (i) simple random sampling (SRS), (ii) balanced sampling, (iii \& iv) the proposed adaptive and oracle sampling designs, respectively, determined by the mean score method for the discrete-time survival analysis. We took equal proportions for the pilot and adaptive samples. The target parameter for the mean score design was the interaction between unfavorable histology and late stage disease. Mean squared error and its bias-variance decomposition are estimated from $1000$ phase two subsamples of $n=400$ from the reduced full cohort ($N=3757$). Reference parameters estimates are from the full cohort analysis using the continuous-time Cox model with complete data on all subjects.}
    \label{table-wilms-cox-N3757}
    \small
    \begin{tabular}{cc rrrrr}
	\hline
	\multicolumn{1}{c}{\multirow{2}{*}{Sampling}}	&	\multirow{2}{*}{Criterion}	&	\multicolumn{5}{c}{Estimation performance by regressor}\\
	\cline{3-7}
		&	&	UH$^1$	&	Stage$^2$	&	Age$^3$	&	dTmr$^4$	&	U$\ast$S$^5$\\
	\hline	
	\multicolumn{1}{c}{Full cohort analysis }	&	Ref.	&	1.027	&	0.292	&	0.064	&	0.022	&	0.620\\
	\hline
	 \multirow{3}{*}{SRS}		
		&	$\sqrt{\textrm{MSE}}$	&	0.388 & 0.313 & 0.046 & 0.033 & 0.600 \\
		&	Bias					&	-0.018 & 0.014 & -0.002 & 0.001 & 0.028 \\
		&	$\sqrt{\textrm{Var}}$		&	0.387 & 0.312 & 0.046 & 0.033 & 0.599 \\ 
	\hline
	\multirow{3}{*}{Balanced}		
		&	$\sqrt{\textrm{MSE}}$	&	0.413 & 0.421 & 0.061 & 0.043 & 0.622 \\
		&	Bias					&	0.057 & 0.020 & 0.010 & 0.006 & -0.008 \\ 
		&	$\sqrt{\textrm{Var}}$		&	0.409 & 0.420 & 0.060 & 0.042 & 0.622 \\
	\hline
	 \multirow{3}{*}{Adaptive}		
		&	$\sqrt{\textrm{MSE}}$	&	0.308 & 0.297 & 0.048 & 0.030 & 0.461 \\ 
		&	Bias					&	-0.000 & -0.006 & 0.007 & 0.003 & 0.039 \\
		&	$\sqrt{\textrm{Var}}$		&	0.308 & 0.296 & 0.048 & 0.029 & 0.459 \\ 
	\hline
	 \multirow{3}{*}{Oracle}		
		&	$\sqrt{\textrm{MSE}}$	&	0.313 & 0.332 & 0.046 & 0.031 & 0.477 \\
		&	Bias					&	-0.001 & 0.001 & 0.000 & 0.000 & 0.035 \\ 
		&	$\sqrt{\textrm{Var}}$		&	0.313 & 0.332 & 0.046 & 0.031 & 0.476 \\ 
	\hline 
	\end{tabular}
    \end{center}
    \vspace{-0.75cm}
    \footnotesize  
    \begin{flushleft}
    	\hspace{1in}$^1$ Unfavorable histology versus favorable; $^2$ disease stage III/IV versus I/II;\\
	    \hspace{1in}$^3$ year at diagnosis; $^4$ tumor diameter (cm); $^5$ interaction effect between UH and Stage.
    \end{flushleft}
\end{table}

Table \ref{table-wilms-cox-N3757} demonstrates performance comparison of different sampling methods when they were applied to the two-phase analysis of the usual Cox model based on 1000 repeated phase two samples in the NWTS data. Unlike the previous analysis in Table \ref{table-wilms-n400} or Figure \ref{figure2}, we assumed that the continuous relapse times were observed for the full cohort in the phase one study and employed the previously developed optimal and adaptive mean score design for the discrete-time survival model \eqref{phz-model} only to design the allocation of the phase two sample. That is, we first calculated the sampling probabilities and associated inverse probability weights for the phase two sample, and then the design-based estimates of the continuous-time Cox model were reported in Table \ref{table-wilms-cox-N3757}. We assumed efficient estimation of the interaction effect between unfavorable histology and disease stage was of primary interest in the adaptive and oracle sampling designs. We used the \texttt{survival} package in R \citep{therneau2019survival} and applied the inverse probability weights of the phase two sample with the weights option of the \texttt{coxph} function.  As shown in Table \ref{table-wilms-cox-N3757},  the adaptive sampling design for the discrete-time analysis also provided efficiency gains for the continuous-time analysis. For example, the variance reduction was about $26\%$ compared to both the simple random sampling and the balanced design. In this example, the adaptive design happened to slightly outperform the oracle one, but we note that these two allocations were for the optimal design for the discrete-time and not the continuous survival model \eqref{phz-model}. We found that the proposed adaptive sampling design provided $20\%$--$25\%$ of efficiency gains in each case varying the target parameter to be (a) unfavorable histology (UH), (b) late stage of the disease, (c) age of diagnosis (year) or (d) tumor diameter (cm) (data not shown).

    \subsection{Design considerations for general censoring patterns} \label{design-censoring}

The proposed method in Section 3 can be applied without modification for the setting where there is random intermittent censoring, as well as censoring at the end of the study; the only difference being that there will be more strata at the intermittent event times. The mean score estimation depends on nonparametric estimation of the probability distribution for the phase two variables conditional on each discrete value of the phase one surrogate. We note that $\hat{\pi}(Y_i, \Delta_i, \mathbf{Z}_i)$ in Section \ref{subsec-mean score} is the empirical estimate of the sampling probability of the $i$-th individual selected into the validation subset, which may suffer from the curse of dimensionality as the number of phase one strata $(Y_i, \Delta_i, \mathbf{Z}_i)$ increases. One could consider an increasing number of discretized intervals to approximate the continuous-time points, but the number of unique continuous-time survival outcomes observed will increase as the sample size increases. For this reason, in our data example we first studied the two-phase analyses of the discrete-time survival models in the previous sections where individuals were right-censored only at the sixth time point, an induced end of follow-up, which is equivalent to a fixed Type I censoring when individuals who were intermittently censored were excluded from analysis. Even in this simple setting in Section \ref{subsec-sim}, there were $(7 \times 4)$-strata for the phase two sampling, resulting from a discrete surrogate $Z$ having four categories. If we further considered random right-censoring for this setting, we would have to estimate sampling probabilities for $(6 \times 2 \times 4)$-strata. The larger number of associated nuisance parameters in this case for MS-A estimator may require larger phase two samples to achieve the expected efficiency gains due to unstable nuisance parameter estimation in small strata. Indeed, in our simulation study with Type I censoring only, we saw that when $n=200$, the MS-A did not outperform MS-SRS. 

We suggest a simple strategy for the proposed adaptive mean score design aimed at under-sampling less informative strata in the pilot study, which may preserve efficiency gains by providing more precise nuisance parameters for the more informative strata. For example, under the random right-censoring assumption, individuals censored before the end of the follow-up period should generally be be less informative than a non-censored individuals with similar covariates for that same period. For a fixed phase two sample size, such under-sampling may enable us to re-allocate the pilot sample for the MS-A so that relatively more informative groups can be up-weighted. In fact, we anticipate that the optimal adaptive design would allocate very few individuals in these strata, so we seek to avoid putting too many individuals into these strata in the pilot. Supplementary Material Tables \ref{table-wilms-cens-n400} and \ref{table-wilms-cox-N3915} provide analogous results for the discrete-time analysis in Table \ref{table-wilms-n400} and continuous time analysis of Table \ref{table-wilms-cox-N3757} for the NWTS, but we analyzed the full cohort ($N=3915$) and under-sampled the censored individuals not relapsing before the end of the follow-up period in the balanced and the pilot samples. Specifically, we allocated a small number of the pilot sample size for individuals censored before the end of the follow-up period (i.e., $y<3$ and $\delta=0$) at each level of histology, and the remaining allocation was equally distributed to the other strata in the pilot sample. In this particular example, we set the under-sampling size to 4, which was approximately half of the balanced sampling size for each stratum in the pilot study with $n=400$. Indeed, we found that the naive balanced sampling of all strata in the pilot often over-sampled censored-groups compared to the oracle design (data not shown), and consequently this simple remedy made the modified adaptive sampling design be closer to the oracle design than did the balanced pilot sample with the proposed adaptive design. As seen in Supplemental Table \ref{table-wilms-cens-n400}, the modified approach produced a $14\%$ variance reduction compared to both the fully balanced and the MS-A with a balanced pilot sample. Additionally, the precision for all estimates, relative to the analogous estimates in Tables \ref{table-wilms-n400} and \ref{table-wilms-cox-N3757}, benefited from including the  intermittently censored $158$ ($=3915-3757$) individuals who had been excluded in Section \ref{subsec-data}, with more gains seen for the MS-BAL and MS-A estimators.  We hypothesize that this modification will allow for more robust efficiency gains for the proposed adaptive design in other settings with intermittent censoring. Supportive simulation studies that examine the  performance of MS-A given the anticipated phase two sample size and other study parameters may provide useful insights to guide refinements of this strategy for a given setting.

\section{Discussion}
	
The mean score method is a practical approach for two-phase studies that allows for a relatively straightforward derivation of an optimal design for a phase two study, one that can minimize the variance of a target parameter given a fixed phase two sample size \citep{reilly1995mean}. In this study, we extended the mean score estimation method for the two-phase analysis of discrete-time survival outcomes. We also derived an adaptive sampling design approach that first draws a pilot phase two sample in order to estimate the nuisance parameters necessary to derive the optimal sampling proportions, similar to the approach of \cite{mcisaac2015adaptive} for binary outcomes. 

Through numerical studies with simulated data and the National Wilms Tumor Study data, we found that the proposed mean score estimator with an adaptive sampling design provided efficiency gains over the complete case estimator, as well as the mean score estimator with simple random or balanced stratified sampling, for selection of the phase two sample.  For the studied settings, as the phase two sample size increased, the proposed adaptive sampling design not only outperformed the simple random sampling and balanced designs consistently but also behaved very close to the oracle design, which depends on the true (generally unknown) population parameters. 

The mean score adaptive optimal design also provided efficiency gains for the two-phase analysis of the continuous-time outcome using the usual Cox proportional hazards model. This design offers a practical and straightforward approach two improve the efficiency of the two-phase estimator for the continuous survival time setting and can be applied for settings with both type I and random right censoring.

There are some limitations of our proposed method. The adaptive design requires estimating nuisance parameters to estimate the strata-specific sampling probabilities that minimize the variance of the target parameter, whose number increases with the number of strata. Many small strata could lead to instability in the necessary estimated nuisance parameters, which in turn could threaten the efficiency of the design. For continuous surrogates, a kernel smoothing approach proposed by \cite{chatterjee2007semiparametric} may be useful. In the case of the Wilms Tumor data, undersampling the early censored individuals in the pilot was an advantageous approach that was compatible with the subsequent derived optimal sample, in that it avoided oversampling uninformative strata, and improved the performance of the adaptive design. Formalization of this strategy needs further study and is a subject for future work.  Incorporation of prior information regarding the sampling priorities for certain strata may be an additional way to improve the performance of the adaptive design, particularly in the case of a small phase two sample. \cite{chenAbstract19} considers a method for incorporation of prior information into multi-wave sampling in a regression framework. 

Two-phase studies are used in a variety of settings. In the era where the analysis of error-prone electronic health records data are of increasing interest,  a phase two sample in which data can be validated to understand the error structure is a critical step towards valid inference.  The proposed method and two-phase study design offer a practical and easy to implement framework for the common setting of survival outcomes, in which both the validated and error-prone exposures can be efficiently combined so that analyses are adjusted for errors in the surrogate data. Future work is needed to expand this method to also handle settings where there is error in both the exposure and survival outcome.

\begin{spacing}{1}
        \bibliography{ref-bib}
        \bibliographystyle{ims}
\end{spacing}

\clearpage

\begin{center}
    \Large
    \textbf{Supplementary Material for ``Two-phase analysis and study design for survival models with error-prone exposures''}
    
    \bigskip
    \large
    Kyunghee Han$^1$, Thomas Lumley$^2$, Bryan E. Shepherd$^3$ and Pamela A. Shaw$^1$
    
    \bigskip
    \normalsize
    $^1$University of Pennsylvania, USA\\
    $^2$University of Auckland, New Zealand\\
    $^3$Vanderbilt University, USA
\end{center}

\renewcommand\thesection{\Alph{section}}
\renewcommand\thesubsection{\Alph{section}.\arabic{subsection}}
 \renewcommand{\thetable}{\Alph{section}.\arabic{table}}
 \renewcommand{\thefigure}{\Alph{section}.\arabic{figure}}

\setcounter{page}{1}
\setcounter{section}{0}
 \setcounter{table}{0}
 \setcounter{figure}{0}

	\section{Technical details}
	
	\subsection{Regularity conditions for Theorem \ref{asymp-normal}} \label{conditions}

We assume there exists $J \geq 1$ such that $P(Y \leq t_J)$ with probability one. Let $\boldsymbol{\theta}_0 \in \Theta_0$ be the true parameter for some open $\Theta_0 \subset \mathbf{R}^{J+d}$ and further assume the following conditions:
\begin{itemize}
	\item[A1.] $ \mathrm{E} |L_1(\boldsymbol{\theta}; Y, \Delta, \mathbf{X})| < \infty$ at $\boldsymbol{\theta}_0$.
	\item[A2.] $L_1(\boldsymbol{\theta}; Y, \Delta, \mathbf{X})$ is three-times continuously differentiable and bounded away from 0 near $\boldsymbol{\theta}_0$ (a.s.).
	\item[A3.] $I_V = - \mathrm{E} \big[\frac{\partial^2}{\partial\boldsymbol{\theta}\partial\boldsymbol{\theta}^\top} \log L_1(\boldsymbol{\theta}; Y, \Delta, \mathbf{X}) \big]$ is positive definite at $\boldsymbol{\theta}_0$.
	\item[A4.] $\pi(y,\delta,\mathbf{z}) > 0$ for all $(y,\delta,\mathbf{z})$.
\end{itemize}
Since we have formulated discrete-time survival models where the censoring time $C$ is bounded away from infinity, the right-censored survival time $Y$ has an upper-bound $t_J$ almost surely so that finite numbers of baseline hazards suffice to specify likelihood for the discrete-time survival model \eqref{phz-model}. Without loss of generality, we write $\boldsymbol{\alpha} = (\alpha_1, \ldots, \alpha_J)^\top$ by the transformation of baseline hazards. Then, the above assumptions A1--A4 are the regularity conditions for the maximum likelihood estimation of finite-dimensional parameters $\boldsymbol{\theta} = (\boldsymbol{\alpha}, \boldsymbol{\beta})$, and therefore, the proof follows Theorem 1 in \cite{reilly1995mean} which we refer the reader to for details.

	\subsection{Some derivations for numerical implementation}\label{technical-details}

For the implementation of the optimal design in our numerical simulations, we empirically estimated certain parameters by generating a large independent sample of $N_0$ individuals. Let $\mathcal{X}_{N_0}^\star = \{ (Y_i^\star, \Delta_i^\star, \mathbf{X}_i^\star, \mathbf{Z}_i^\star) : 1 \leq i \leq N_0 \}$ be an external random sample of $(Y, \Delta, \mathbf{X}, \mathbf{Z})$ independent on $\mathcal{X}_N$ or $\mathcal{X}_{\textrm{I},N}$. We estimate unknown quantities in the optimal sampling design \eqref{optimal-design} by
\begin{eqnarray}
\begin{split}
	\widehat{I}_V^\star 
	    &= -\frac{1}{N_0}\sum_{i=1}^{N_0} \frac{\partial^2}{\partial \boldsymbol{\theta} \partial \boldsymbol{\theta}^\top} \log L_1(\boldsymbol{\theta}; Y_i^\star, \Delta_i^\star, \mathbf{X}_i^\star),\\
    \widehat{\mathrm{Var}}^\star \big(U_1(\boldsymbol{\theta}) | y, \delta, z \big)
        &= \frac{N^\star(y,\delta,\mathbf{z})}{N^\star(y,\delta,\mathbf{z})-1} \big\{ \hat{\mu}_2^\star(\boldsymbol{\theta}; y,\delta,\mathbf{z}) - \hat{\mu}_1^\star(\boldsymbol{\theta}; y,\delta,\mathbf{z})^2\big\},
\end{split}\label{est-external-1}
\end{eqnarray}
where $N^\star(y,\delta,\mathbf{z})= \sum_{i=1}^{N_0} \mathbb{I}(Y_i^\star=y, \Delta_i^\star=\delta, \mathbf{Z}_i^\star = \mathbf{z})$ and
\begin{align}
    \hat{\mu}_\ell^\star(\boldsymbol{\theta}; y,\delta,\mathbf{z}) = \frac{1}{N^\star(y,\delta,\mathbf{z})} \sum_{i=1}^{N_0} U_1(\boldsymbol{\theta} ; Y_i^\star, \Delta_i^\star, \mathbf{X}_i^\star)^\ell \cdot \mathbb{I}(Y_i^\star=y, \Delta_i^\star=\delta, \mathbf{Z}_i^\star = \mathbf{z})\label{est-external-2}
\end{align}
for $\ell=1,2$.

We now introduce some derivations for Newton-Raphson algorithm. Recall that the discrete-time survival model \eqref{phz-model} under the odds transformation $g_1(u) = \frac{u}{1-u}$ is given by $\textrm{logit}(\lambda_j(\mathbf{x}))  =  \alpha_j+ \boldsymbol{\beta}^\top \mathbf{x}$, where $\alpha_j = \textrm{logit}(\lambda_{0j})$ is the logit transformation of the baseline hazard. The maximum likelihood estimates of $\boldsymbol{\alpha}$ and $\boldsymbol{\beta}$ are obtained by solving score equations of \eqref{em-re} with
\begin{eqnarray}
\begin{split}
	\frac{\partial}{\partial \alpha_j}{Q}_{N}(\boldsymbol{\alpha}, \boldsymbol{\beta}) 
		&= \sum_{i \in \mathcal{I}} \hat{\pi}(Y_i, \Delta_i, \mathbf{Z}_i)^{-1} \mathbb{I}(j \leq J(i)) \left( {D}_{ij} - \frac{e^{ \alpha_j + \boldsymbol{\beta}^\top \mathbf{X}_i}}{1 +  e^{ \alpha_j + \boldsymbol{\beta}^\top \mathbf{X}_i}} \right) = 0,\\
	\frac{\partial}{\partial \boldsymbol{\beta}}{Q}_{N}(\boldsymbol{\alpha}, \boldsymbol{\beta})
		&= \sum_{i \in \mathcal{I}} \hat{\pi}(Y_i, \Delta_i, \mathbf{Z}_i)^{-1}  \sum_{j=1}^{J(i)} \left( {D}_{ij} -  \frac{e^{ \alpha_j + \boldsymbol{\beta}^\top \mathbf{X}_i}}{1 +  e^{ \alpha_j + \boldsymbol{\beta}^\top \mathbf{X}_i}} \right) \mathbf{X}_i = \mathbf{0}.
\end{split}\label{score-logit}
\end{eqnarray}
We numerically solve the system of equations \eqref{score-logit} with the Newton-Raphson algorithm using the associated Hessian matrix whose components consist of
\begin{eqnarray}
\begin{split}
	\frac{\partial^2}{\partial \alpha_j \partial \alpha_{j'}}{Q}_{N}(\boldsymbol{\alpha}, \boldsymbol{\beta}) 
		&= -\sum_{i \in \mathcal{I}}  \hat{\pi}(Y_i, \Delta_i, \mathbf{Z}_i)^{-1}  \mathbb{I}(j \leq J(i)) \frac{e^{ \alpha_j + \boldsymbol{\beta}^\top \mathbf{X}_i}}{\big(1 +  e^{ \alpha_j + \boldsymbol{\beta}^\top \mathbf{X}_i}\big)^2} \, \mathbb{I}(j=j'),\\
	\frac{\partial^2}{\partial \alpha_j \partial \boldsymbol{\beta}}{Q}_{N}(\boldsymbol{\alpha}, \boldsymbol{\beta}) 
		&= -\sum_{i \in \mathcal{I}} \hat{\pi}(Y_i, \Delta_i, \mathbf{Z}_i)^{-1} \mathbb{I}(j \leq J(i))\frac{e^{ \alpha_j + \boldsymbol{\beta}^\top \mathbf{X}_i}}{\big(1 +  e^{ \alpha_j + \boldsymbol{\beta}^\top \mathbf{X}_i}\big)^2}  \, \mathbf{X}_i,\\
	\frac{\partial^2}{\partial \boldsymbol{\beta} \partial \boldsymbol{\beta}^\top}{Q}_{N}(\boldsymbol{\alpha}, \boldsymbol{\beta}) 
		&= -\sum_{i \in \mathcal{I}} \hat{\pi}(Y_i, \Delta_i, \mathbf{Z}_i)^{-1} \sum_{j=1}^{J(i)}\frac{e^{ \alpha_j + \boldsymbol{\beta}^\top \mathbf{X}_i}}{\big(1 +  e^{ \alpha_j + \boldsymbol{\beta}^\top \mathbf{X}_i}\big)^2} \, \mathbf{X}_i \mathbf{X}_i^\top.
\end{split}\label{hessian-logit}
\end{eqnarray}

Similarly, suppose the complementary log transformation $g_2(u) = -\log(1-u)$ defines the true survival model \eqref{phz-model}. Then it can be easily seen that $\textrm{logit}(\lambda_j(\mathbf{x}))  =  \exp\big(e^{\alpha_j+ \boldsymbol{\beta}^\top \mathbf{x}}\big) -1$, where $\alpha_j = \log(-\log(1-\lambda_{0j}))$ is the complementary log-log transformation of the baseline hazard. The maximum likelihood estimates of $\boldsymbol{\theta} = (\boldsymbol{\alpha}, \boldsymbol{\beta})$ are obtained by solving score equations of \eqref{em-re} with
\begin{eqnarray}
\begin{split}
	\frac{\partial}{\partial \alpha_j}{Q}_{N}(\boldsymbol{\alpha}, \boldsymbol{\beta}) 
		&= \sum_{i \in \mathcal{I}} \hat{\pi}(Y_i, \Delta_i, \mathbf{Z}_i)^{-1} \mathbb{I}(j \leq J(i))\left( {D}_{ij} \frac{e^{ \alpha_j + \boldsymbol{\beta}^\top \mathbf{X}_i}}{1-e^{\exp(\alpha_j + \boldsymbol{\beta}^\top \mathbf{X}_i)}} - e^{ \alpha_j + \boldsymbol{\beta}^\top \mathbf{X}_i} \right) = 0,\\
	\frac{\partial}{\partial \boldsymbol{\beta}}{Q}_{N}(\boldsymbol{\alpha}, \boldsymbol{\beta}) 
		&= \sum_{i \in \mathcal{I}} \hat{\pi}(Y_i, \Delta_i, \mathbf{Z}_i)^{-1}  \sum_{j=1}^{J(i)} \left( {D}_{ij} \frac{e^{ \alpha_j + \boldsymbol{\beta}^\top \mathbf{X}_i}}{1-e^{\exp(\alpha_j + \boldsymbol{\beta}^\top \mathbf{X}_i)}} - e^{ \alpha_j + \boldsymbol{\beta}^\top \mathbf{X}_i} \right) \mathbf{X}_i = \mathbf{0}.
\end{split}\label{score-cloglog}
\end{eqnarray}
We numerically solve the system of equations \eqref{score-cloglog} with the Newton-Raphson algorithm using the associated Hessian matrix whose components consist of
\begin{eqnarray}
\begin{split}
	\frac{\partial^2}{\partial \alpha_j \partial \alpha_{j'}}{Q}_{N}(\boldsymbol{\alpha}, \boldsymbol{\beta}) 
		&= -\sum_{i \in \mathcal{I}}  \hat{\pi}(Y_i, \Delta_i, \mathbf{Z}_i)^{-1} \mathbb{I}(j \leq J(i)) \left\{ {D}_{ij} \frac{e^{ \alpha_j + \boldsymbol{\beta}^\top \mathbf{X}_i}}{1-e^{\exp(\alpha_j + \boldsymbol{\beta}^\top \mathbf{X}_i)}} \right. \times\\
		&\qquad\qquad\qquad \left. \bigg( 1 - \frac{e^{ \alpha_j + \boldsymbol{\beta}^\top \mathbf{X}_i}e^{-\exp(\alpha_j + \boldsymbol{\beta}^\top \mathbf{X}_i)}}{1-e^{\exp(\alpha_j + \boldsymbol{\beta}^\top \mathbf{X}_i)}} \bigg) - e^{ \alpha_j + \boldsymbol{\beta}^\top \mathbf{X}_i} \right\}\, \mathbb{I}(j=j'),\\
	\frac{\partial^2}{\partial \alpha_j \partial \boldsymbol{\beta}}{Q}_{N}(\boldsymbol{\alpha}, \boldsymbol{\beta}) 
		&= -\sum_{i \in \mathcal{I}}  \hat{\pi}(Y_i, \Delta_i, \mathbf{Z}_i)^{-1} \mathbb{I}(j \leq J(i)) \left\{ {D}_{ij} \frac{e^{ \alpha_j + \boldsymbol{\beta}^\top \mathbf{X}_i}}{1-e^{\exp(\alpha_j + \boldsymbol{\beta}^\top \mathbf{X}_i)}} \right. \times\\ 
		&\qquad\qquad\qquad \left. \bigg( 1 - \frac{e^{ \alpha_j + \boldsymbol{\beta}^\top \mathbf{X}_i}e^{-\exp(\alpha_j + \boldsymbol{\beta}^\top \mathbf{X}_i)}}{1-e^{\exp(\alpha_j + \boldsymbol{\beta}^\top \mathbf{X}_i)}} \bigg) - e^{ \alpha_j + \boldsymbol{\beta}^\top \mathbf{X}_i} \right\}\, \mathbf{X}_i,\\
	\frac{\partial^2}{\partial \boldsymbol{\beta} \partial \boldsymbol{\beta}^\top}{Q}_{N}(\boldsymbol{\alpha}, \boldsymbol{\beta}) 
		&= -\sum_{i \in \mathcal{I}}  \hat{\pi}(Y_i, \Delta_i, \mathbf{Z}_i)^{-1} \sum_{j=1}^{J(i)}\left\{ {D}_{ij} \frac{e^{ \alpha_j + \boldsymbol{\beta}^\top \mathbf{X}_i}}{1-e^{\exp(\alpha_j + \boldsymbol{\beta}^\top \mathbf{X}_i)}} \right. \times\\
		&\qquad\qquad\qquad \left. \bigg( 1 - \frac{e^{ \alpha_j + \boldsymbol{\beta}^\top \mathbf{X}_i}e^{-\exp(\alpha_j + \boldsymbol{\beta}^\top \mathbf{X}_i)}}{1-e^{\exp(\alpha_j + \boldsymbol{\beta}^\top \mathbf{X}_i)}} \bigg) - e^{ \alpha_j + \boldsymbol{\beta}^\top \mathbf{X}_i} \right\}\, \mathbf{X}_i \mathbf{X}_i^\top.
\end{split}\label{hessian-cloglog}
\end{eqnarray}

\newpage

\section{Additional numerical results}

\subsection{Supplemental tables for the simulation study}
	
\bigskip
\begin{table}[!htbp]
    \begin{center}
    \caption{Relative performance for the estimation of $\beta_1$ is compared for (i) the complete case analysis with simple random sampling (CC-SRS), (ii) the mean score method with simple random sampling (MS-SRS), (iii) a design-based estimation with balanced sample, equivalent to the mean score (MS-BAL), (iv \& v) the mean score estimation with adaptive sampling (MS-A) and the optimal sampling design (MS-O), for varying sample sizes. Results for the full cohort estimator based on complete data are provided as a benchmark.  Mean squared error (MSE) and its bias-variance decomposition are estimated from $1000$ Monte Carlo replications, where the censoring rate was $30\%$. The adaptive and optimal sampling designs were for efficient estimation of $X_1$ with $\beta_1=\log(1.5) \approx 0.405$. In all scenarios, we took equal proportions for the pilot and adaptive samples.}
    \label{table-sim-cens30}
    \small
    \begin{tabular}{ccc  rrr c rrr }
	\hline
	\multicolumn{1}{c}{\multirow{3}{*}{Sampling}}	& 	\multicolumn{1}{c}{\multirow{3}{*}{Estimation}}	&	\multirow{3}{*}{Criterion}	&	\multicolumn{7}{c}{Estimation performance by sample sizes}\\
	\cline{4-10}
		&	&	&	\multicolumn{3}{c}{\scriptsize{$N=4000$}}	& &	\multicolumn{3}{c}{\footnotesize{$n=400$}}\\
	\cline{4-6} \cline{8-10}
		&	&	&	\footnotesize{$n=200$}	&	\footnotesize{$n=400$}	&	\footnotesize{$n=800$}	& &	\scriptsize{$N=2000$}	&	\scriptsize{$N=4000$}	&	\scriptsize{$N=8000$}\\
	\hline
	\multirow{3}{*}{Full cohort}		
			&	\multirow{3}{*}{CC}
						&	$\sqrt{\textrm{MSE}}$	&	0.079	&	0.079	&	0.079	& &		0.112 	&	0.079	&	0.056	\\
		\multicolumn{2}{c}{}	&	Bias					&	0.004	&	0.004	&	0.004	& &		0.000	&	0.004	&	0.001	\\
		\multicolumn{2}{c}{}	&	$\sqrt{\textrm{Var}}$		&	0.079	&	0.079	&	0.079	& &		0.112	&	0.079	&	0.056	\\
	\hline
	 \multirow{6}{*}{SRS}		
			&	\multirow{3}{*}{CC}
						&	$\sqrt{\textrm{MSE}}$	&	0.408	&	0.284	&	0.198	& &		0.281 	&	0.284	&	0.282	\\
		\multicolumn{2}{c}{}	&	Bias					&	0.013	&	0.012	&	0.011	& &		0.013	&	0.012	&	0.015	\\
		\multicolumn{2}{c}{}	&	$\sqrt{\textrm{Var}}$		&	0.408	&	0.284	&	0.197	& &		0.281	&	0.284	&	0.282	\\
	\cline{2-10}
		&	\multirow{3}{*}{MS}
						&	$\sqrt{\textrm{MSE}}$	&	0.268	&	0.177	&	0.132	& &		0.191 	&	0.177	&	0.170	\\
		\multicolumn{2}{c}{}	&	Bias					&	0.067	&	0.016	&	0.012	& &		0.014	&	0.016	&	0.017	\\
		\multicolumn{2}{c}{}	&	$\sqrt{\textrm{Var}}$		&	0.259	&	0.176	&	0.131	& &		0.190 	&	0.176	&	0.169	\\
	\hline
	\multirow{3}{*}{Balanced}		
			&	\multirow{3}{*}{MS}
						&	$\sqrt{\textrm{MSE}}$	&	0.302	&	0.216	&	0.153	& &		0.219 	&	0.216	&	0.202	\\
		\multicolumn{2}{c}{}	&	Bias					&	0.023	&	0.017	&	0.015	& &		0.010	&	0.017	&	0.018	\\
		\multicolumn{2}{c}{}	&	$\sqrt{\textrm{Var}}$		&	0.301	&	0.215	&	0.152	& &		0.219 	&	0.215	&	0.201	\\
	\hline
	 \multirow{3}{*}{Adaptive}		
			&	\multirow{3}{*}{MS}
						&	$\sqrt{\textrm{MSE}}$	&	0.279	&	0.168	&	0.114	& &		0.181 	&	0.168	&	0.159	\\
		\multicolumn{2}{c}{}	&	Bias					&	0.038	&	0.024	&	0.011	& &		0.012	&	0.024	&	0.017	\\
		\multicolumn{2}{c}{}	&	$\sqrt{\textrm{Var}}$		&	0.276	&	0.166	&	0.113	& &		0.181	&	0.166	&	0.158	\\
	\hline
	\multirow{3}{*}{Oracle}		
			&	\multirow{3}{*}{MS}
						&	$\sqrt{\textrm{MSE}}$	&	0.213	&	0.155	&	0.112	& &		0.166 	&	0.155	&	0.137	\\
		\multicolumn{2}{c}{}	&	Bias					&	0.025	&	0.001	&	0.001	& &		0.005	&	0.001	&	0.002	\\
		\multicolumn{2}{c}{}	&	$\sqrt{\textrm{Var}}$		&	0.212	&	0.155	&	0.112	& &		0.166	&	0.155	&	0.137	\\
	\hline 
	\end{tabular}
    \end{center}
\end{table}

\begin{table}[!htbp]
    \begin{center}
    \caption{Relative performance for the estimation of $\beta_1$ is compared for (i) the complete case analysis with simple random sampling (CC-SRS), (ii) the mean score method with simple random sampling (MS-SRS), (iii) a design-based estimation with balanced sample, equivalent to the mean score (MS-BAL), (iv \& v) the mean score estimation with adaptive sampling (MS-A) and the optimal sampling design (MS-O), for varying sample sizes. Results for the full cohort estimator based on complete data are provided as a benchmark.  Mean squared error (MSE) and its bias-variance decomposition are estimated from $1000$ Monte Carlo replications, where the censoring rate was $70\%$. The adaptive and optimal sampling designs were for efficient estimation of $X_1$ with $\beta_1=\log(1.5) \approx 0.405$. In all scenarios, we took equal proportions for the pilot and adaptive samples.}
    \label{table-sim-cens70}
    \small
    \begin{tabular}{ccc  rrr c rrr }
	\hline
	\multicolumn{1}{c}{\multirow{3}{*}{Sampling}}	& 	\multicolumn{1}{c}{\multirow{3}{*}{Estimation}}	&	\multirow{3}{*}{Criterion}	&	\multicolumn{7}{c}{Estimation performance by sample sizes}\\
	\cline{4-10}
		&	&	&	\multicolumn{3}{c}{\scriptsize{$N=4000$}}	& &	\multicolumn{3}{c}{\footnotesize{$n=400$}}\\
	\cline{4-6} \cline{8-10}
		&	&	&	\footnotesize{$n=200$}	&	\footnotesize{$n=400$}	&	\footnotesize{$n=800$}	& &	\scriptsize{$N=2000$}	&	\scriptsize{$N=4000$}	&	\scriptsize{$N=8000$}\\
	\hline
	\multirow{3}{*}{Full cohort}		
			&	\multirow{3}{*}{CC}
						&	$\sqrt{\textrm{MSE}}$	&	0.119	&	0.119	&	0.119	& &		0.175 	&	0.119	&	0.086	\\
		\multicolumn{2}{c}{}	&	Bias					&	0.000	&	0.000	&	0.000	& &		0.004	&	0.000	&	0.003	\\
		\multicolumn{2}{c}{}	&	$\sqrt{\textrm{Var}}$		&	0.119	&	0.119	&	0.119	& &		0.175	&	0.119	&	0.086	\\
	\hline
	 \multirow{6}{*}{SRS}		
			&	\multirow{3}{*}{CC}
						&	$\sqrt{\textrm{MSE}}$	&	0.630	&	0.420	&	0.299	& &		0.442 	&	0.420	&	0.419	\\
		\multicolumn{2}{c}{}	&	Bias					&	0.025	&	0.016	&	0.002	& &		-0.005	&	0.016	&	0.018	\\
		\multicolumn{2}{c}{}	&	$\sqrt{\textrm{Var}}$		&	0.630	&	0.420	&	0.299	& &		0.442	&	0.420	&	0.418	\\
	\cline{2-10}
		&	\multirow{3}{*}{MS}
						&	$\sqrt{\textrm{MSE}}$	&	0.477	&	0.304	&	0.208	& &		0.328	&	0.304	&	0.287	\\
		\multicolumn{2}{c}{}	&	Bias					&	0.177	&	0.078	&	0.018	& &		0.065	&	0.078	&	0.081	\\
		\multicolumn{2}{c}{}	&	$\sqrt{\textrm{Var}}$		&	0.443	&	0.294	&	0.207	& &		0.322 	&	0.294	&	0.273	\\
	\hline
	\multirow{3}{*}{Balanced}		
			&	\multirow{3}{*}{MS}
						&	$\sqrt{\textrm{MSE}}$	&	0.535	&	0.330	&	0.225	& &		0.347 	&	0.330	&	0.327	\\
		\multicolumn{2}{c}{}	&	Bias					&	0.092	&	0.026	&	0.005	& &		0.033	&	0.026	&	0.029	\\
		\multicolumn{2}{c}{}	&	$\sqrt{\textrm{Var}}$		&	0.527	&	0.329	&	0.224	& &		0.346	&	0.329	&	0.326	\\
	\hline
	 \multirow{3}{*}{Adaptive}		
			&	\multirow{3}{*}{MS}
						&	$\sqrt{\textrm{MSE}}$	&	0.411	&	0.250	&	0.178	& &		0.259 	&	0.250	&	0.225	\\
		\multicolumn{2}{c}{}	&	Bias					&	0.018	&	0.007	&	0.001	& &		0.003	&	0.007	&	0.003	\\
		\multicolumn{2}{c}{}	&	$\sqrt{\textrm{Var}}$		&	0.410	&	0.250	&	0.178	& &		0.259	&	0.250	&	0.225	\\
	\hline
	\multirow{3}{*}{Oracle}		
			&	\multirow{3}{*}{MS}
						&	$\sqrt{\textrm{MSE}}$	&	0.315	&	0.215	&	0.174	& &		0.250 	&	0.215	&	0.224	\\
		\multicolumn{2}{c}{}	&	Bias					&	0.023	&	0.010	&	0.000	& &		0.017	&	0.010	&	0.008	\\
		\multicolumn{2}{c}{}	&	$\sqrt{\textrm{Var}}$		&	0.314	&	0.214	&	0.174	& &		0.249	&	0.214	&	0.224	\\
	\hline 
	\end{tabular}
    \end{center}
\end{table}

\begin{table}[!htbp]
    \begin{center}
    \caption{Relative performance for the estimation of regression parameters is compared for (i) the complete case analysis with simple random sampling (CC-SRS), (ii) the mean score method with simple random sampling (MS-SRS), (iii) a design-based estimation with balanced sample, equivalent to the mean score (MS-BAL), (iv \& v) the mean score estimation with adaptive sampling (MS-A) and the optimal sampling design (MS-O), where $n=200$ subsample was drawn from the full cohort size of $N=4000$. Results for the full cohort estimator based on complete data are provided as a benchmark.  Mean squared error (MSE) and its bias-variance decomposition are estimated from $1000$ Monte Carlo replications, where the censoring rate was $50\%$. The adaptive and optimal sampling designs were for efficient estimation of $X_1$ with $\beta_1=\log(1.5) \approx 0.405$.}
    \label{table-sim-app-n200-N4000}
    \small
    \resizebox{\columnwidth}{!}{%
    \begin{tabular}{cccc rrrrrr rrrr}
	\hline
	\multicolumn{1}{c}{\multirow{2}{*}{Sample size}}	&	\multicolumn{1}{c}{\multirow{2}{*}{Sampling}}	& 	\multicolumn{1}{c}{\multirow{2}{*}{Estimation}}	&	\multirow{2}{*}{Criterion}	&	\multicolumn{10}{c}{Estimation performance by regression coefficient}\\
	\cline{5-14}
		&	&	&	&	$\alpha_1$	&	$\alpha_2$	&	$\alpha_3$	&	$\alpha_4$	&	$\alpha_5$	&	$\alpha_6$	&	$\beta_1$	&	$\beta_2$	&	$\beta_3$	&	$\beta_4$\\
	\hline	
	\multirow{13}{*}{$n=200$}	
		&	 \multirow{6}{*}{SRS}		
			&	\multirow{3}{*}{CC}
						&	$\sqrt{\textrm{MSE}}$	&	0.547 & 0.520 & 0.447 & 0.429 & 0.409 & 0.396	&	0.470 & 0.555 & 0.214 & 0.203 \\
		\multicolumn{3}{c}{}	&	Bias					&	-0.086 & -0.076 & -0.064 & -0.034 & -0.033 & -0.008	&	0.017 & -0.017 & 0.013 & -0.009 \\
		\multicolumn{3}{c}{}	&	$\sqrt{\textrm{Var}}$		&	0.540 & 0.514 & 0.443 & 0.428 & 0.408 & 0.396	&	0.470 & 0.555 & 0.214 & 0.202 \\
	\cline{3-14}
	\multirow{12}{*}{$N=4000$}	
		&	&	\multirow{3}{*}{MS}
						&	$\sqrt{\textrm{MSE}}$	&	0.479 & 0.381 & 0.330 & 0.316 & 0.306 & 0.301	&	0.332 & 0.602 & 0.236 & 0.219 \\
		\multicolumn{3}{c}{}	&	Bias					&	-0.267 & -0.166 & -0.112 & -0.086 & -0.074 & -0.056	&	0.053 & -0.021 & 0.020 & -0.017 \\
		\multicolumn{3}{c}{}	&	$\sqrt{\textrm{Var}}$		&	0.398 & 0.343 & 0.310 & 0.304 & 0.297 & 0.295	&	0.330 & 0.601 & 0.235 & 0.218 \\
	\cline{2-14}
		&	 \multirow{3}{*}{Balanced}		
			&	\multirow{3}{*}{MS}
						&	$\sqrt{\textrm{MSE}}$	&	0.405 & 0.401 & 0.399 & 0.398 & 0.398 & 0.405	&	0.400 & 0.859 & 0.317 & 0.299 \\
		\multicolumn{3}{c}{}	&	Bias					&	-0.023 & -0.011 & -0.009 & -0.002 & 0.009 & 0.034	&	0.042 & -0.034 & 0.025 & -0.017 \\
		\multicolumn{3}{c}{}	&	$\sqrt{\textrm{Var}}$		&	0.404 & 0.400 & 0.399 & 0.398 & 0.398 & 0.403	&	0.398 & 0.858 & 0.317 & 0.298 \\
	\cline{2-14}
		&	 \multirow{3}{*}{Adaptive}		
			&	\multirow{3}{*}{MS}
						&	$\sqrt{\textrm{MSE}}$	&	0.332 & 0.324 & 0.323 & 0.323 & 0.325 & 0.326	&	0.374 & 0.720 & 0.255 & 0.251 \\
		\multicolumn{3}{c}{}	&	Bias					&	-0.018 & -0.007 & -0.006 & -0.001 & 0.006 & 0.024	&	0.049 & -0.065 & 0.025 & -0.028 \\
		\multicolumn{3}{c}{}	&	$\sqrt{\textrm{Var}}$		&	0.331 & 0.324 & 0.323 & 0.323 & 0.325 & 0.325	&	0.371 & 0.717 & 0.254 & 0.249 \\
	\cline{2-14}
		&	 \multirow{3}{*}{Oracle}		
			&	\multirow{3}{*}{MS}
						&	$\sqrt{\textrm{MSE}}$	&	0.258 & 0.253 & 0.255 & 0.252 & 0.251 & 0.254	&	0.253 & 0.519 & 0.198 & 0.189 \\
		\multicolumn{3}{c}{}	&	Bias					&	-0.013 & 0.003 & 0.002 & 0.005 & 0.007 & 0.018	&	0.009 & -0.027 & 0.001 & -0.017 \\
		\multicolumn{3}{c}{}	&	$\sqrt{\textrm{Var}}$		&	0.257 & 0.253 & 0.255 & 0.252 & 0.251 & 0.253 	&	0.252 & 0.518 & 0.198 & 0.188 \\
	\hline 
	\multicolumn{3}{c}{\multirow{3}{*}{Full cohort analysis}}
						&	$\sqrt{\textrm{MSE}}$	&	0.107 & 0.096 & 0.091 & 0.085 & 0.080 & 0.079	&	0.094 & 0.116 & 0.044 & 0.042 \\ 
		\multicolumn{3}{c}{}	&	Bias					&	-0.010 & 0.000 & -0.003 & -0.002 & -0.005 & -0.002	&	0.003 & -0.001 & -0.000 & 0.000 \\
		\multicolumn{3}{c}{}	&	$\sqrt{\textrm{Var}}$		&	0.107 & 0.096 & 0.091 & 0.085 & 0.080 & 0.079	&	0.094 & 0.116 & 0.044 & 0.042 \\
	\hline
	\end{tabular}
	}
    \end{center}
    \vspace{-0.75cm}
    \footnotesize  
    \begin{flushleft}
    	$(\alpha_1, \alpha_2, \alpha_3, \alpha_4, \alpha_5, \alpha_6) = (-3.410, -3.027, -2.641, -2.249, -1.849, -1.435)$;\\
    	$(\beta_1, \beta_2, \beta_3, \beta_4) = (\log(1.5), \log(0.7), \log(1.3), -\log(1.3)) \approx (0.405, -0.357, 0.262, -0.262)$.
    \end{flushleft}
\end{table}

\begin{table}[!htbp]
    \begin{center}
    \caption{Relative performance for the estimation of regression parameters is compared for (i) the complete case analysis with simple random sampling (CC-SRS), (ii) the mean score method with simple random sampling (MS-SRS), (iii) a design-based estimation with balanced sample, equivalent to the mean score (MS-BAL), (iv \& v) the mean score estimation with adaptive sampling (MS-A) and the optimal sampling design (MS-O), where $n=400$ subsample was drawn from the full cohort size of $N=4000$. Results for the full cohort estimator based on complete data are provided as a benchmark.  Mean squared error (MSE) and its bias-variance decomposition are estimated from $1000$ Monte Carlo replications, where the censoring rate was $50\%$. The adaptive and optimal sampling designs were for efficient estimation of $X_1$ with $\beta_1=\log(1.5) \approx 0.405$.}
    \label{table-sim-app-n400-N4000}
    \small
    \resizebox{\columnwidth}{!}{%
    \begin{tabular}{cccc rrrrrr rrrr}
	\hline
	\multicolumn{1}{c}{\multirow{2}{*}{Sample size}}	&	\multicolumn{1}{c}{\multirow{2}{*}{Sampling}}	& 	\multicolumn{1}{c}{\multirow{2}{*}{Estimation}}	&	\multirow{2}{*}{Criterion}	&	\multicolumn{10}{c}{Estimation performance by regression coefficient}\\
	\cline{5-14}
		&	&	&	&	$\alpha_1$	&	$\alpha_2$	&	$\alpha_3$	&	$\alpha_4$	&	$\alpha_5$	&	$\alpha_6$	&	$\beta_1$	&	$\beta_2$	&	$\beta_3$	&	$\beta_4$\\
	\hline	
	\multirow{13}{*}{$n=400$}	
		&	 \multirow{6}{*}{SRS}		
			&	\multirow{3}{*}{CC}
						&	$\sqrt{\textrm{MSE}}$	&	0.377 & 0.347 & 0.319 & 0.295 & 0.288 & 0.280	&	0.330 & 0.404 & 0.146 & 0.143 \\
		\multicolumn{3}{c}{}	&	Bias					&	-0.050 & -0.045 & -0.037 & -0.014 & -0.025 & -0.008	&	0.014 & 0.000 & 0.006 & -0.004 \\
		\multicolumn{3}{c}{}	&	$\sqrt{\textrm{Var}}$		&	0.374 & 0.344 & 0.316 & 0.295 & 0.287 & 0.280	&	0.329 & 0.404 & 0.146 & 0.143 \\
	\cline{3-14}
	\multirow{12}{*}{$N=4000$}	
		&	&	\multirow{3}{*}{MS}
						&	$\sqrt{\textrm{MSE}}$	&	0.253 & 0.226 & 0.216 & 0.209 & 0.204 & 0.203	&	0.220 & 0.423 & 0.154 & 0.147 \\
		\multicolumn{3}{c}{}	&	Bias					&	-0.081 & -0.046 & -0.039 & -0.032 & -0.030 & -0.022	&	0.035 & 0.005 & 0.007 & -0.005 \\
		\multicolumn{3}{c}{}	&	$\sqrt{\textrm{Var}}$		&	0.240 & 0.222 & 0.213 & 0.207 & 0.202 & 0.202	&	0.217 & 0.423 & 0.154 & 0.147 \\
	\cline{2-14}
		&	 \multirow{3}{*}{Balanced}		
			&	\multirow{3}{*}{MS}
						&	$\sqrt{\textrm{MSE}}$	&	0.284 & 0.279 & 0.276 & 0.276 & 0.277 & 0.277	&	0.278 & 0.582 & 0.200 & 0.204 \\
		\multicolumn{3}{c}{}	&	Bias					&	-0.014 & -0.003 & -0.004 & -0.000 & 0.003 & 0.016	&	0.024 & -0.007 & 0.006 & -0.015 \\
		\multicolumn{3}{c}{}	&	$\sqrt{\textrm{Var}}$		&	0.284 & 0.279 & 0.276 & 0.276 & 0.277 & 0.277	&	0.277 & 0.582 & 0.200 & 0.204 \\
	\cline{2-14}
		&	 \multirow{3}{*}{Adaptive}		
			&	\multirow{3}{*}{MS}
						&	$\sqrt{\textrm{MSE}}$	&	0.210 & 0.200 & 0.201 & 0.201 & 0.198 & 0.201	&	0.197 & 0.426 & 0.147 & 0.147 \\
		\multicolumn{3}{c}{}	&	Bias					&	-0.013 & -0.003 & -0.005 & -0.003 & -0.002 & 0.005	&	0.007 & -0.014 & 0.006 & -0.009 \\
		\multicolumn{3}{c}{}	&	$\sqrt{\textrm{Var}}$		&	0.210 & 0.200 & 0.200 & 0.201 & 0.198 & 0.200	&	0.197 & 0.426 & 0.147 & 0.147 \\
	\cline{2-14}
		&	 \multirow{3}{*}{Oracle}		
			&	\multirow{3}{*}{MS}
						&	$\sqrt{\textrm{MSE}}$	&	0.189 & 0.184 & 0.182 & 0.177 & 0.178 & 0.178	&	0.182 & 0.354 & 0.134 & 0.132 \\
		\multicolumn{3}{c}{}	&	Bias					&	-0.003 & 0.007 & 0.005 & 0.006 & 0.005 & 0.011	&	0.003 & -0.014 & -0.005 & -0.008 \\
		\multicolumn{3}{c}{}	&	$\sqrt{\textrm{Var}}$		&	0.189 & 0.184 & 0.182 & 0.177 & 0.178 & 0.177	&	0.182 & 0.354 & 0.133 & 0.132 \\
	\hline 
	\multicolumn{3}{c}{\multirow{3}{*}{Full cohort analysis}}
						&	$\sqrt{\textrm{MSE}}$	&	0.107 & 0.096 & 0.091 & 0.085 & 0.080 & 0.079	&	0.094 & 0.116 & 0.044 & 0.042 \\ 
		\multicolumn{3}{c}{}	&	Bias					&	-0.010 & 0.000 & -0.003 & -0.002 & -0.005 & -0.002	&	0.003 & -0.001 & -0.000 & 0.000 \\
		\multicolumn{3}{c}{}	&	$\sqrt{\textrm{Var}}$		&	0.107 & 0.096 & 0.091 & 0.085 & 0.080 & 0.079	&	0.094 & 0.116 & 0.044 & 0.042 \\
	\hline
	\end{tabular}
	}
    \end{center}
    \vspace{-0.75cm}
    \footnotesize  
    \begin{flushleft}
    	$(\alpha_1, \alpha_2, \alpha_3, \alpha_4, \alpha_5, \alpha_6) = (-3.410, -3.027, -2.641, -2.249, -1.849, -1.435)$;\\
    	$(\beta_1, \beta_2, \beta_3, \beta_4) = (\log(1.5), \log(0.7), \log(1.3), -\log(1.3)) \approx (0.405, -0.357, 0.262, -0.262)$.
    \end{flushleft}
\end{table}

\begin{table}[!htbp]
    \begin{center}
    \caption{Relative performance for the estimation of regression parameters is compared for (i) the complete case analysis with simple random sampling (CC-SRS), (ii) the mean score method with simple random sampling (MS-SRS), (iii) a design-based estimation with balanced sample, equivalent to the mean score (MS-BAL), (iv \& v) the mean score estimation with adaptive sampling (MS-A) and the optimal sampling design (MS-O), where $n=800$ subsample was drawn from the full cohort size of $N=4000$. Results for the full cohort estimator based on complete data are provided as a benchmark.  Mean squared error (MSE) and its bias-variance decomposition are estimated from $1000$ Monte Carlo replications, where the censoring rate was $50\%$. The adaptive and optimal sampling designs were for efficient estimation of $X_1$ with $\beta_1=\log(1.5) \approx 0.405$.}
    \label{table-sim-app-n800-N4000}
    \small
    \resizebox{\columnwidth}{!}{%
    \begin{tabular}{cccc rrrrrr rrrr}
	\hline
	\multicolumn{1}{c}{\multirow{2}{*}{Sample size}}	&	\multicolumn{1}{c}{\multirow{2}{*}{Sampling}}	& 	\multicolumn{1}{c}{\multirow{2}{*}{Estimation}}	&	\multirow{2}{*}{Criterion}	&	\multicolumn{10}{c}{Estimation performance by regression coefficient}\\
	\cline{5-14}
		&	&	&	&	$\alpha_1$	&	$\alpha_2$	&	$\alpha_3$	&	$\alpha_4$	&	$\alpha_5$	&	$\alpha_6$	&	$\beta_1$	&	$\beta_2$	&	$\beta_3$	&	$\beta_4$\\
	\hline	
	\multirow{13}{*}{$n=800$}	
		&	 \multirow{6}{*}{SRS}		
			&	\multirow{3}{*}{CC}
						&	$\sqrt{\textrm{MSE}}$	&	0.262 & 0.239 & 0.220 & 0.206 & 0.199 & 0.191	&	0.228 & 0.286 & 0.101 & 0.101 \\ 
		\multicolumn{3}{c}{}	&	Bias					&	-0.030 & -0.018 & -0.016 & -0.007 & -0.016 & -0.005	&	0.005 & 0.002 & 0.005 & -0.004 \\ 
		\multicolumn{3}{c}{}	&	$\sqrt{\textrm{Var}}$		&	0.260 & 0.238 & 0.220 & 0.206 & 0.198 & 0.191	&	0.228 & 0.286 & 0.101 & 0.101 \\
	\cline{3-14}
	\multirow{12}{*}{$N=4000$}	
		&	&	\multirow{3}{*}{MS}
						&	$\sqrt{\textrm{MSE}}$	&	0.169 & 0.156 & 0.151 & 0.149 & 0.146 & 0.144	&	0.155 & 0.292 & 0.104 & 0.102 \\
		\multicolumn{3}{c}{}	&	Bias					&	-0.022 & -0.008 & -0.009 & -0.008 & -0.009 & -0.004	&	0.004 & 0.001 & 0.007 & -0.004 \\
		\multicolumn{3}{c}{}	&	$\sqrt{\textrm{Var}}$		&	0.167 & 0.156 & 0.151 & 0.148 & 0.145 & 0.144	&	0.155 & 0.292 & 0.103 & 0.102 \\
	\cline{2-14}
		&	 \multirow{3}{*}{Balanced}		
			&	\multirow{3}{*}{MS}
						&	$\sqrt{\textrm{MSE}}$	&	0.202 & 0.196 & 0.193 & 0.191 & 0.189 & 0.190	&	0.194 & 0.396 & 0.145 & 0.141 \\
		\multicolumn{3}{c}{}	&	Bias					&	-0.017 & -0.006 & -0.009 & -0.007 & -0.007 & 0.000	&	0.010 & 0.008 & -0.002 & 0.002 \\
		\multicolumn{3}{c}{}	&	$\sqrt{\textrm{Var}}$		&	0.201 & 0.195 & 0.193 & 0.190 & 0.189 & 0.190	&	0.194 & 0.396 & 0.145 & 0.141 \\
	\cline{2-14}
		&	 \multirow{3}{*}{Adaptive}		
			&	\multirow{3}{*}{MS}
						&	$\sqrt{\textrm{MSE}}$	&	0.154 & 0.149 & 0.145 & 0.142 & 0.139 & 0.140	&	0.147 & 0.280 & 0.099 & 0.102 \\
		\multicolumn{3}{c}{}	&	Bias					&	-0.010 & -0.001 & -0.003 & -0.003 & -0.003 & 0.001	&	0.006 & -0.010 & 0.005 & -0.003 \\
		\multicolumn{3}{c}{}	&	$\sqrt{\textrm{Var}}$		&	0.154 & 0.149 & 0.145 & 0.142 & 0.139 & 0.140	&	0.147 & 0.280 & 0.099 & 0.102 \\
	\cline{2-14}
		&	 \multirow{3}{*}{Oracle}		
			&	\multirow{3}{*}{MS}
						&	$\sqrt{\textrm{MSE}}$	&	0.147 & 0.139 & 0.137 & 0.135 & 0.131 & 0.131	&	0.133 & 0.261 & 0.096 & 0.095 \\
		\multicolumn{3}{c}{}	&	Bias					&	-0.004 & 0.006 & 0.003 & 0.004 & 0.003 & 0.007	&	0.005 & -0.018 & 0.002 & -0.005 \\
		\multicolumn{3}{c}{}	&	$\sqrt{\textrm{Var}}$		&	0.147 & 0.139 & 0.137 & 0.135 & 0.131 & 0.131	&	0.133 & 0.260 & 0.096 & 0.095 \\
	\hline 
	\multicolumn{3}{c}{\multirow{3}{*}{Full cohort analysis}}
						&	$\sqrt{\textrm{MSE}}$	&	0.107 & 0.096 & 0.091 & 0.085 & 0.080 & 0.079	&	0.094 & 0.116 & 0.044 & 0.042 \\ 
		\multicolumn{3}{c}{}	&	Bias					&	-0.010 & 0.000 & -0.003 & -0.002 & -0.005 & -0.002	&	0.003 & -0.001 & -0.000 & 0.000 \\
		\multicolumn{3}{c}{}	&	$\sqrt{\textrm{Var}}$		&	0.107 & 0.096 & 0.091 & 0.085 & 0.080 & 0.079	&	0.094 & 0.116 & 0.044 & 0.042 \\
	\hline
	\end{tabular}
	}
    \end{center}
    \vspace{-0.75cm}
    \footnotesize  
    \begin{flushleft}
    	$(\alpha_1, \alpha_2, \alpha_3, \alpha_4, \alpha_5, \alpha_6) = (-3.410, -3.027, -2.641, -2.249, -1.849, -1.435)$;\\
    	$(\beta_1, \beta_2, \beta_3, \beta_4) = (\log(1.5), \log(0.7), \log(1.3), -\log(1.3)) \approx (0.405, -0.357, 0.262, -0.262)$.
    \end{flushleft}
\end{table}

\begin{table}[!htbp]
    \begin{center}
    \caption{Relative performance for the estimation of regression parameters is compared for (i) the complete case analysis with simple random sampling (CC-SRS), (ii) the mean score method with simple random sampling (MS-SRS), (iii) a design-based estimation with balanced sample, equivalent to the mean score (MS-BAL), (iv \& v) the mean score estimation with adaptive sampling (MS-A) and the optimal sampling design (MS-O), where $n=400$ subsample was drawn from the full cohort size of $N=2000$. Results for the full cohort estimator based on complete data are provided as a benchmark.  Mean squared error (MSE) and its bias-variance decomposition are estimated from $1000$ Monte Carlo replications, where the censoring rate was $50\%$. The adaptive and optimal sampling designs were for efficient estimation of $X_1$ with $\beta_1=\log(1.5) \approx 0.405$.}
    \label{table-sim-app-n400-N2000}
    \small
    \resizebox{\columnwidth}{!}{%
    \begin{tabular}{cccc rrrrrr rrrr}
	\hline
	\multicolumn{1}{c}{\multirow{2}{*}{Sample size}}	&	\multicolumn{1}{c}{\multirow{2}{*}{Sampling}}	& 	\multicolumn{1}{c}{\multirow{2}{*}{Estimation}}	&	\multirow{2}{*}{Criterion}	&	\multicolumn{10}{c}{Estimation performance by regression coefficient}\\
	\cline{5-14}
		&	&	&	&	$\alpha_1$	&	$\alpha_2$	&	$\alpha_3$	&	$\alpha_4$	&	$\alpha_5$	&	$\alpha_6$	&	$\beta_1$	&	$\beta_2$	&	$\beta_3$	&	$\beta_4$\\
	\hline
	\multirow{13}{*}{$n=400$}	
		&	 \multirow{6}{*}{SRS}		
			&	\multirow{3}{*}{CC}
						&	$\sqrt{\textrm{MSE}}$	&	0.381 & 0.342 & 0.323 & 0.288 & 0.281 & 0.284	&	0.324 & 0.419 & 0.145 & 0.147 \\
		\multicolumn{3}{c}{}	&	Bias					&	-0.049 & -0.044 & -0.039 & -0.025 & -0.020 & -0.019	&	-0.011 & 0.029 & 0.006 & -0.005 \\
		\multicolumn{3}{c}{}	&	$\sqrt{\textrm{Var}}$		&	0.378 & 0.340 & 0.321 & 0.287 & 0.280 & 0.283	&	0.324 & 0.418 & 0.145 & 0.147 \\ 
	\cline{3-14}
	\multirow{12}{*}{$N=2000$}	
		&	&	\multirow{3}{*}{MS}
						&	$\sqrt{\textrm{MSE}}$	&	0.267 & 0.245 & 0.232 & 0.224 & 0.217 & 0.223	&	0.237 & 0.434 & 0.150 & 0.156 \\
		\multicolumn{3}{c}{}	&	Bias					&	-0.072 & -0.050 & -0.044 & -0.036 & -0.028 & -0.031	&	0.020 & 0.027 & 0.005 & -0.004 \\
		\multicolumn{3}{c}{}	&	$\sqrt{\textrm{Var}}$		&	0.258 & 0.240 & 0.228 & 0.221 & 0.215 & 0.221	&	0.236 & 0.433 & 0.150 & 0.156  \\
	\cline{2-14}
		&	 \multirow{3}{*}{Balanced}		
			&	\multirow{3}{*}{MS}
						&	$\sqrt{\textrm{MSE}}$	&	0.278 & 0.275 & 0.267 & 0.266 & 0.263 & 0.266	&	0.276 & 0.550 & 0.200 & 0.198 \\
		\multicolumn{3}{c}{}	&	Bias					&	-0.024 & -0.018 & -0.019 & -0.014 & -0.006 & -0.003	&	0.026 & -0.005 & 0.013 & -0.005 \\
		\multicolumn{3}{c}{}	&	$\sqrt{\textrm{Var}}$		&	0.277 & 0.274 & 0.266 & 0.265 & 0.263 & 0.266	&	0.275 & 0.550 & 0.199 & 0.198 \\
	\cline{2-14}
		&	 \multirow{3}{*}{Adaptive}		
			&	\multirow{3}{*}{MS}
						&	$\sqrt{\textrm{MSE}}$	&	0.233 & 0.220 & 0.211 & 0.210 & 0.207 & 0.210	&	0.214 & 0.421 & 0.152 & 0.148 \\ 
		\multicolumn{3}{c}{}	&	Bias					&	-0.019 & -0.014 & -0.016 & -0.012 & -0.005 & -0.007	&	0.004 & 0.006 & 0.010 & -0.016 \\ 
		\multicolumn{3}{c}{}	&	$\sqrt{\textrm{Var}}$		&	0.232 & 0.219 & 0.210 & 0.209 & 0.207 & 0.210	&	0.214 & 0.421 & 0.152 & 0.147 \\
	\cline{2-14}
		&	 \multirow{3}{*}{Oracle}		
			&	\multirow{3}{*}{MS}
						&	$\sqrt{\textrm{MSE}}$	&	0.215 & 0.204 & 0.195 & 0.195 & 0.186 & 0.195	&	0.202 & 0.383 & 0.134 & 0.136 \\
		\multicolumn{3}{c}{}	&	Bias					&	-0.021 & -0.012 & -0.014 & -0.010 & -0.004 & -0.008	&	-0.012 & 0.021 & 0.001 & -0.004 \\
		\multicolumn{3}{c}{}	&	$\sqrt{\textrm{Var}}$		&	0.214 & 0.203 & 0.194 & 0.194 & 0.186 & 0.195	&	0.202 & 0.382 & 0.134 & 0.136 \\
	\hline 
	\multicolumn{3}{c}{\multirow{3}{*}{Full cohort analysis}}
						&	$\sqrt{\textrm{MSE}}$	&	0.152 & 0.136 & 0.124 & 0.121 & 0.112 & 0.118		&	0.129 & 0.172 & 0.062 & 0.061 \\
		\multicolumn{3}{c}{}	&	Bias					&	-0.009 & -0.005 & -0.008 & -0.006 & -0.002 & -0.008	&	-0.004 & 0.010 & 0.001 & -0.001 \\
		\multicolumn{3}{c}{}	&	$\sqrt{\textrm{Var}}$		&	0.152 & 0.136 & 0.124 & 0.121 & 0.112 & 0.118		&	0.129 & 0.171 & 0.062 & 0.061 \\
	\hline
	\end{tabular}
	}
    \end{center}
    \vspace{-0.75cm}
    \footnotesize  
    \begin{flushleft}
    	$(\alpha_1, \alpha_2, \alpha_3, \alpha_4, \alpha_5, \alpha_6) = (-3.410, -3.027, -2.641, -2.249, -1.849, -1.435)$;\\
    	$(\beta_1, \beta_2, \beta_3, \beta_4) = (\log(1.5), \log(0.7), \log(1.3), -\log(1.3)) \approx (0.405, -0.357, 0.262, -0.262)$.
    \end{flushleft}
\end{table}

\begin{table}[!htbp]
    \begin{center}
    \caption{Relative performance for the estimation of regression parameters is compared for (i) the complete case analysis with simple random sampling (CC-SRS), (ii) the mean score method with simple random sampling (MS-SRS), (iii) a design-based estimation with balanced sample, equivalent to the mean score (MS-BAL), (iv \& v) the mean score estimation with adaptive sampling (MS-A) and the optimal sampling design (MS-O), where $n=400$ subsample was drawn from the full cohort size of $N=8000$. Results for the full cohort estimator based on complete data are provided as a benchmark.  Mean squared error (MSE) and its bias-variance decomposition are estimated from $1000$ Monte Carlo replications, where the censoring rate was $50\%$. The adaptive and optimal sampling designs were for efficient estimation of $X_1$ with $\beta_1=\log(1.5) \approx 0.405$.}
    \label{table-sim-app-n400-N8000}
    \small
    \resizebox{\columnwidth}{!}{%
    \begin{tabular}{cccc rrrrrr rrrr}
	\hline
	\multicolumn{1}{c}{\multirow{2}{*}{Sample size}}	&	\multicolumn{1}{c}{\multirow{2}{*}{Sampling}}	& 	\multicolumn{1}{c}{\multirow{2}{*}{Estimation}}	&	\multirow{2}{*}{Criterion}	&	\multicolumn{10}{c}{Estimation performance by regression coefficient}\\
	\cline{5-14}
		&	&	&	&	$\alpha_1$	&	$\alpha_2$	&	$\alpha_3$	&	$\alpha_4$	&	$\alpha_5$	&	$\alpha_6$	&	$\beta_1$	&	$\beta_2$	&	$\beta_3$	&	$\beta_4$\\
	\hline	
	\multirow{13}{*}{$n=400$}	
		&	 \multirow{6}{*}{SRS}		
			&	\multirow{3}{*}{CC}
						&	$\sqrt{\textrm{MSE}}$	&	0.381 & 0.334 & 0.316 & 0.289 & 0.281 & 0.280	&	0.321 & 0.419 & 0.155 & 0.149 \\ 
		\multicolumn{3}{c}{}	&	Bias					&	-0.042 & -0.037 & -0.034 & -0.020 & -0.014 & -0.004	&	-0.001 & 0.001 & 0.012 & -0.007 \\
		\multicolumn{3}{c}{}	&	$\sqrt{\textrm{Var}}$		&	0.379 & 0.332 & 0.314 & 0.288 & 0.280 & 0.280	&	0.321 & 0.419 & 0.154 & 0.149 \\ 
	\cline{3-14}
	\multirow{12}{*}{$N=8000$}	
		&	&	\multirow{3}{*}{MS}
						&	$\sqrt{\textrm{MSE}}$	&	0.236 & 0.211 & 0.208 & 0.201 & 0.199 & 0.200	&	0.200 & 0.435 & 0.162 & 0.157 \\ 
		\multicolumn{3}{c}{}	&	Bias					&	-0.071 & -0.043 & -0.034 & -0.028 & -0.022 & -0.017	&	0.027 & -0.000 & 0.015 & -0.009 \\ 
		\multicolumn{3}{c}{}	&	$\sqrt{\textrm{Var}}$		&	0.225 & 0.206 & 0.205 & 0.199 & 0.198 & 0.200	&	0.198 & 0.435 & 0.161 & 0.157 \\ 
	\cline{2-14}
		&	 \multirow{3}{*}{Balanced}		
			&	\multirow{3}{*}{MS}
						&	$\sqrt{\textrm{MSE}}$	&	0.272 & 0.271 & 0.271 & 0.268 & 0.269 & 0.270	&	0.265 & 0.572 & 0.211 & 0.207 \\
		\multicolumn{3}{c}{}	&	Bias					&	-0.012 & -0.009 & -0.007 & -0.004 & 0.004 & 0.014	&	0.021 & -0.002 & 0.007 & -0.017 \\
		\multicolumn{3}{c}{}	&	$\sqrt{\textrm{Var}}$		&	0.272 & 0.271 & 0.271 & 0.268 & 0.269 & 0.270	&	0.264 & 0.572 & 0.211 & 0.207 \\
	\cline{2-14}
		&	 \multirow{3}{*}{Adaptive}		
			&	\multirow{3}{*}{MS}
						&	$\sqrt{\textrm{MSE}}$	&	0.196 & 0.194 & 0.195 & 0.193 & 0.194 & 0.195	&	0.190 & 0.417 & 0.153 & 0.148 \\ 
		\multicolumn{3}{c}{}	&	Bias					&	-0.009 & -0.008 & -0.007 & -0.005 & -0.001 & 0.004	&	0.002 & -0.014 & 0.008 & 0.001 \\
		\multicolumn{3}{c}{}	&	$\sqrt{\textrm{Var}}$		&	0.196 & 0.193 & 0.195 & 0.193 & 0.194 & 0.195	&	0.189 & 0.417 & 0.153 & 0.148 \\
	\cline{2-14}
		&	 \multirow{3}{*}{Oracle}		
			&	\multirow{3}{*}{MS}
						&	$\sqrt{\textrm{MSE}}$	&	0.178 & 0.174 & 0.170 & 0.170 & 0.171 & 0.172	&	0.174 & 0.362 & 0.134 & 0.133 \\ 
		\multicolumn{3}{c}{}	&	Bias					&	-0.006 & -0.004 & -0.002 & -0.002 & 0.002 & 0.005	&	-0.005 & -0.010 & -0.001 & 0.008 \\
		\multicolumn{3}{c}{}	&	$\sqrt{\textrm{Var}}$		&	0.178 & 0.174 & 0.170 & 0.170 & 0.171 & 0.172	&	0.174 & 0.362 & 0.134 & 0.133 \\
	\hline 
	\multicolumn{3}{c}{\multirow{3}{*}{Full cohort analysis}}
						&	$\sqrt{\textrm{MSE}}$	&	0.076 & 0.071 & 0.063 & 0.060 & 0.057 & 0.055	&	0.067 & 0.085 & 0.030 & 0.031 \\ 
		\multicolumn{3}{c}{}	&	Bias					&	-0.002 & -0.001 & -0.001 & -0.001 & 0.001 & 0.001	&	0.000 & -0.003 & 0.001 & 0.000 \\
		\multicolumn{3}{c}{}	&	$\sqrt{\textrm{Var}}$		&	0.076 & 0.071 & 0.063 & 0.060 & 0.057 & 0.055	&	0.067 & 0.085 & 0.030 & 0.031 \\
	\hline
	\end{tabular}
	}
    \end{center}
    \vspace{-0.75cm}
    \footnotesize  
    \begin{flushleft}
    	$(\alpha_1, \alpha_2, \alpha_3, \alpha_4, \alpha_5, \alpha_6) = (-3.410, -3.027, -2.641, -2.249, -1.849, -1.435)$;\\
    	$(\beta_1, \beta_2, \beta_3, \beta_4) = (\log(1.5), \log(0.7), \log(1.3), -\log(1.3)) \approx (0.405, -0.357, 0.262, -0.262)$.
    \end{flushleft}
\end{table}

\clearpage

\subsection{Two-phase analysis with modified sampling design in the NWTS data example}\label{appendix-nwts}

\begin{table}[!htbp]
    \begin{center}
    \caption{Modification of the proposed mean score sampling design for the discrete-time survival analysis of the National Wilms Tumor Study to include all individuals regardless of the time to censoring ($N=3915$). Similarly to Table \ref{table-wilms-n400}, we compare five different methods (CC-SRS, MS-SRS, MS-BAL, MS-A, MS-O) as well as two modified designs by under-sampling the censored individuals not relapsing before the end of the follow-up period in the balanced and the pilot sample$^{\dagger,\ddagger}$. The optimal sampling design was estimated using the full cohort data. The MS-A and MS-O designs are for efficient estimation of the interaction effect between unfavorable histology and disease stage.  Results from the full cohort analysis with complete data are presented as a benchmark. Mean squared error and its bias-variance decomposition are estimated using $1000$ phase two samples of $n=400$ from the full cohort ($N=3915$).}
    \label{table-wilms-cens-n400}
    \small
    \resizebox{\columnwidth}{!}{%
    \begin{tabular}{ccc rrrrrr c rrrrr}
	\hline
	\multicolumn{1}{c}{\multirow{2}{*}{Sampling}}	& 	\multicolumn{1}{c}{\multirow{2}{*}{Estimation}}	&	\multirow{2}{*}{Criterion}	&	\multicolumn{6}{c}{Baseline hazard in complementary log-log scale}		&	&	\multicolumn{5}{c}{Regression coefficient}\\
	\cline{4-9}\cline{11-15}
		&	&	&	0.5yr	&	1yr	&	1.5yr	&	2yr	&	2.5yr	&	3yr	&	&	UH$^1$	&	Stage$^2$	&	Age$^3$	&	dTmr$^4$	&	U$\ast$S$^5$\\
	\hline	
	\multicolumn{2}{c}{Full cohort analysis }	&	Ref.	&	-4.074	&	-3.916	&	-4.373	&	-5.037	&	-5.378	&	-5.737	&	&	1.087	&	0.287	&	0.064	&	0.031	&	0.632\\
	\hline
	 \multirow{6}{*}{SRS}		
			&	\multirow{3}{*}{CC}
						&	$\sqrt{\textrm{MSE}}$	&	0.469 & 0.460 & 0.479 & 0.706 & 1.537 & 2.601	&	&	0.640 & 0.340 & 0.050 & 0.034 & 0.763 \\
		\multicolumn{2}{c}{}	&	Bias					&	-0.074 & -0.067 & -0.074 & -0.136 & -0.283 & -0.694	&	&	-0.025 & -0.011 & -0.001 & 0.001 & 0.070 \\
		\multicolumn{2}{c}{}	&	$\sqrt{\textrm{Var}}$		&	0.463 & 0.455 & 0.473 & 0.693 & 1.511 & 2.507	&	&	0.640 & 0.340 & 0.050 & 0.034 & 0.759 \\
	\cline{2-15}
		&	\multirow{3}{*}{MS}
						&	$\sqrt{\textrm{MSE}}$	&	0.433 & 0.428 & 0.429 & 0.580 & 1.464 & 2.587	&	&	0.635 & 0.355 & 0.051 & 0.037 & 0.815 \\
		\multicolumn{2}{c}{}	&	Bias					&	-0.028 & -0.023 & -0.035 & -0.109 & -0.330 & -0.829	&	&	-0.102 & -0.005 & 0.001 & 0.001 & 0.131 \\
		\multicolumn{2}{c}{}	&	$\sqrt{\textrm{Var}}$		&	0.432 & 0.428 & 0.428 & 0.570 & 1.426 & 2.450	&	&	0.627 & 0.355 & 0.051 & 0.037 & 0.804 \\
	\hline
	\multirow{3}{*}{Balanced}		
			&	\multirow{3}{*}{MS}
						&	$\sqrt{\textrm{MSE}}$	&	0.512 & 0.504 & 0.493 & 0.485 & 0.482 & 0.481	&	&	0.460 & 0.437 & 0.068 & 0.043 & 0.712 \\
		\multicolumn{2}{c}{}	&	Bias					&	-0.123 & -0.105 & -0.088 & -0.077 & -0.071 & -0.067	&	&	0.048 & 0.046 & 0.018 & 0.004 & -0.051 \\
		\multicolumn{2}{c}{}	&	$\sqrt{\textrm{Var}}$		&	0.497 & 0.492 & 0.485 & 0.479 & 0.477 & 0.476	&	&	0.457 & 0.435 & 0.066 & 0.042 & 0.710 \\
	\hline
	\multirow{2}{*}{Balanced}		
			&	\multirow{3}{*}{MS}
						&	$\sqrt{\textrm{MSE}}$	&	0.436 & 0.428 & 0.419 & 0.412 & 0.410 & 0.409	&	&	0.403 & 0.371 & 0.061 & 0.036 & 0.614	 \\
	\multirow{2}{*}{($^\dagger$modified)}	&	&	Bias				&	-0.090 & -0.075 & -0.062 & -0.054 & -0.049 & -0.046	&	&	0.038 & 0.009 & 0.016 & 0.003 & 0.002	 \\
		\multicolumn{2}{c}{}	&	$\sqrt{\textrm{Var}}$		&	0.426 & 0.421 & 0.415 & 0.409 & 0.407 & 0.406	&	&	0.401 & 0.371 & 0.059 & 0.036 & 0.614	 \\
	\hline
	 \multirow{3}{*}{Adaptive}		
			&	\multirow{3}{*}{MS}
						&	$\sqrt{\textrm{MSE}}$	&	0.350 & 0.343 & 0.336 & 0.331 & 0.329 & 0.327	&	&	0.326 & 0.313 & 0.045 & 0.028 & 0.519 \\
		\multicolumn{2}{c}{}	&	Bias					&	-0.071 & -0.060 & -0.052 & -0.047 & -0.044 & -0.042	&	&	-0.010 & 0.002 & 0.009 & 0.003 & 0.056 \\
		\multicolumn{2}{c}{}	&	$\sqrt{\textrm{Var}}$		&	0.342 & 0.337 & 0.332 & 0.327 & 0.326 & 0.325	&	&	0.326 & 0.313 & 0.044 & 0.027 & 0.516 \\
	\hline
	 \multirow{2}{*}{Adaptive}		
			&	\multirow{3}{*}{MS}
						&	$\sqrt{\textrm{MSE}}$	&	0.342 & 0.336 & 0.329 & 0.324 & 0.323 & 0.322	&	&	0.290 & 0.266 & 0.042 & 0.027 & 0.448 \\
	 \multirow{2}{*}{($^\ddagger$modified)}	&	&	Bias				&	-0.060 & -0.052 & -0.045 & -0.040 & -0.038 & -0.036	&	&	-0.003 & -0.014 & 0.009 & 0.002 & 0.057 \\
		\multicolumn{2}{c}{}	&	$\sqrt{\textrm{Var}}$		&	0.337 & 0.332 & 0.326 & 0.322 & 0.320 & 0.320	&	&	0.290 & 0.265 & 0.041 & 0.027 & 0.445 \\
	\hline
	 \multirow{3}{*}{Oracle}		
			&	\multirow{3}{*}{MS}
						&	$\sqrt{\textrm{MSE}}$	&	0.269 & 0.267 & 0.264 & 0.261 & 0.261 & 0.260	&	&	0.234 & 0.232 & 0.035 & 0.022 & 0.379 \\
		\multicolumn{2}{c}{}	&	Bias					&	-0.007 & -0.009 & -0.008 & -0.006 & -0.006 & -0.005	&	&	0.006 & 0.008 & 0.002 & -0.000 & 0.030 \\
		\multicolumn{2}{c}{}	&	$\sqrt{\textrm{Var}}$		&	0.269 & 0.267 & 0.264 & 0.261 & 0.261 & 0.260	&	&	0.234 & 0.232 & 0.034 & 0.022 & 0.377 \\
	\hline 
	\end{tabular}
	}
    \end{center}
    \vspace{-0.75cm}
    \footnotesize  
    \begin{flushleft}
    	$^1$ Unfavorable histology versus favorable; $^2$ disease stage III/IV versus I/II;\\
	$^3$ year at diagnosis; $^4$ tumor diameter (cm); $^5$ interaction effect between UH and Stage.
    \end{flushleft}
\end{table}

\begin{table}[!t]
    \begin{center}
    \caption{Modification of the proposed mean score sampling design for the two-phase continuous-time Cox model analysis of time to relapse in the National Wilms Tumor Study to include all individuals regardless of the time to censoring ($N=3915$). Similarly to Table \ref{table-wilms-cox-N3757}, we used inverse probability weights (IPW) for the two-phase analysis for four different sampling designs for the second phase; (i) simple random sampling (SRS), (ii) balanced sampling, (iii \& iv) the proposed adaptive and oracle sampling designs as well as two modified designs by under-sampling the censored individuals not relapsing before the end of the follow-up period in the balanced and the pilot sample.$^{\dagger,\ddagger}$ Here, the adaptive and oracle designs were determined by the mean score method for the discrete-time survival analysis. The target parameter for the mean score design was the interaction between unfavorable histology and late stage disease. Mean squared error and its bias-variance decomposition are estimated from $1000$ phase two subsamples of $n=400$ from the full cohort. Reference parameters estimates are from the full cohort analysis using the continuous-time Cox model with complete data on all subjects.}
    \label{table-wilms-cox-N3915}
    \small
    \begin{tabular}{cc rrrrr}
	\hline
	\multicolumn{1}{c}{\multirow{2}{*}{Sampling}}	&	\multirow{2}{*}{Criterion}	&	\multicolumn{5}{c}{Estimation performance by regressor}\\
	\cline{3-7}
		&	&	UH$^1$	&	Stage$^2$	&	Age$^3$	&	dTmr$^4$	&	U$\ast$S$^5$\\
	\hline	
	\multicolumn{1}{c}{Full cohort analysis }	&	Ref.	&	1.050	&	0.297	&	0.065	&	0.021	&	0.620\\
	\hline
	 \multirow{3}{*}{SRS}		
		&	$\sqrt{\textrm{MSE}}$	&	0.444 & 0.322 & 0.049 & 0.033 & 0.648 \\
		&	Bias					&	-0.092 & 0.009 & -0.000 & 0.000 & 0.115 \\
		&	$\sqrt{\textrm{Var}}$		&	0.434 & 0.322 & 0.049 & 0.033 & 0.638 \\ 
	\hline
	\multirow{3}{*}{Balanced}		
		&	$\sqrt{\textrm{MSE}}$	&	0.478 & 0.519 & 0.074 & 0.049 & 0.752 \\	
		&	Bias					&	0.061 & 0.032 & 0.016 & 0.006 & -0.037 \\ 
		&	$\sqrt{\textrm{Var}}$		&	0.474 & 0.518 & 0.072 & 0.048 & 0.751 \\
	\hline
	\multirow{2}{*}{Balanced}		
		&	$\sqrt{\textrm{MSE}}$	&	0.440 & 0.434 & 0.065 & 0.043 & 0.643 \\
	\multirow{2}{*}{($^\dagger$modified)}		
		&	Bias					&	0.039 & 0.012 & 0.011 & 0.008 & 0.025 \\ 
		&	$\sqrt{\textrm{Var}}$		&	0.438 & 0.434 & 0.065 & 0.043 & 0.643 \\
	\hline
	\multirow{3}{*}{Adaptive}			
		&	$\sqrt{\textrm{MSE}}$	&	0.338 & 0.373 & 0.052 & 0.034 & 0.541 \\ 	
		&	Bias					&	-0.001 & 0.005 & 0.005 & 0.002 & 0.035 \\
		&	$\sqrt{\textrm{Var}}$		&	0.338 & 0.373 & 0.052 & 0.034 & 0.540 \\ 
	\hline
	\multirow{2}{*}{Adaptive}			
		&	$\sqrt{\textrm{MSE}}$	&	0.328 & 0.325 & 0.050 & 0.031 & 0.493 \\ 
	\multirow{2}{*}{($^\dagger$modified)}		
		&	Bias					&	-0.023 & 0.000 & 0.008 & 0.002 & 0.053 \\
		&	$\sqrt{\textrm{Var}}$		&	0.327 & 0.325 & 0.049 & 0.031 & 0.491 \\ 
	\hline
	 \multirow{3}{*}{Oracle}		
		&	$\sqrt{\textrm{MSE}}$	&	0.255 & 0.250 & 0.041 & 0.025 & 0.396 \\
		&	Bias					&	-0.001 & -0.006 & 0.004 & 0.001 & 0.043 \\ 
		&	$\sqrt{\textrm{Var}}$		&	0.255 & 0.250 & 0.041 & 0.025 & 0.393 \\ 
	\hline 
	\end{tabular}
    \end{center}
    \vspace{-0.75cm}
    \footnotesize  
    \begin{flushleft}
    	\hspace{1in}$^1$ Unfavorable histology versus favorable; $^2$ disease stage III/IV versus I/II;\\
	    \hspace{1in}$^3$ year at diagnosis; $^4$ tumor diameter (cm); $^5$ interaction effect between UH and Stage.
    \end{flushleft}
\end{table}

\end{document}